\documentclass[a4paper,fleqn]{cas-dc}

\usepackage{siunitx}

\usepackage{amsmath}
\usepackage[numbers]{natbib}
\setcitestyle{sort&compress}

\usepackage{appendix}
\usepackage{tabularx}
\usepackage{multirow}
\usepackage{array}
\usepackage{caption}
\usepackage{subcaption}
\usepackage{overpic}
\usepackage{threeparttable}
\usepackage{graphicx}

\usepackage{natbib}

\def\tsc#1{\csdef{#1}{\textsc{\lowercase{#1}}\xspace}}
\tsc{WGM}
\tsc{QE}

\begin{document}
\let\WriteBookmarks\relax
\def\floatpagepagefraction{1}
\def\textpagefraction{.001}
\let\printorcid\relax 

\shorttitle{}    

\shortauthors{Haipeng Li et al.}

\title[mode = title]{The symbolic partition with generalized Koopman analysis}  

\author[1]{Haipeng Li}
\author[1]{Pengfei Guo}
\author[1,2]{Yueheng Lan}
\cormark[1]

\address[1]{School of Science, Beijing University of Posts and Telecommunications, Beijing 100876, China} 
\address[2]{State Key Lab of Information Photonics and Optical Communications, Beijing University of Posts and Telecommunications, Beijing 100876, China}
\cortext[1]{Corresponding author}  

\begin{abstract}	
	Symbolic dynamics serves as a crucial tool in the study of chaotic systems, prompting extensive research into various methods for symbolic partitioning. The limitations of these methods are heuristic and empirical for the partition the multivariate chaotic state space. Notably, the use of operator theory in partitioning the multivariable chaotic series into precise symbolic cells has been underexplored. 
In this paper, we propose a novel symbolic partition method, referred to as Koopman Analysis(KA) method, exploiting Koopman operator theory to address the symbolic partition, especially multivariate chaotic time series. 
We map the chaotic time series into the basis functions to obtain the approximate representation of the Koopman operator.
Then we transpose the Koopman approximate matrix and subsequently perform spectral decomposition to obtain the Koopman left eigenfunctions.
We apply KA method to one-dimensional unimodal chaotic map to obtain Koopman left eigenfunctions. 
Then we find some particular eigenfunctions whose eigenvalues are zero, some of which can be used to identify the symbolic boundary of region composed of chaotic series in that the oscillation coincides with the subregion where the theoretical symbolic boundary is located. We refer to the function as the Valid Left Eigenfunction with Zero(VLEZ). Based on the number of oscillations, we further classify VLEZ into two categories which are both valid for localizing the subregion containing coarse symbolic boundary. Then, for refining the coarse symbolic boundary, we modify the KA method applicable to chaotic localized subregion and further propose the Generalized Koopman Analysis (GKA) method. 
The KA method can be also applied to the multimodal maps, multivariate chaotic maps and hyperchaotic maps and their noisy version. For these complex chaotic series, we take KA method with low resolution and only obtain the characteristic eigenfunctions (CLE) instead of VLEZ. Fortunately, we obtain estimated coarse symbolic boundary and
then we redistribute basis functions based on CLE. Under new basis functions, we take KA method and obtain primary coarse symbolic boundary. Then, we can refine the symbolic boundary via GKA method. 
The precise symbolic boundary is represented correctly and validly via comparing with verified results from previous research.
Our method is applicable to chaotic time series of diverse dimensions and complexities, accommodating noisy chaotic data with relative accuracy. The current work can be well further expand to higher dimensional and more complex time series due to its interpretability and availability. 

\end{abstract}


\maketitle

\section{Introduction\label{sec:intro}}

Practical time series frequently exhibit chaotic phenomena, which pose challenges for detailed analysis. Symbolic dynamics offers a simplified topological description that proves invaluable in navigating the complexities of chaotic system. Introduced by Morse and Hedlund in 1938\cite{1938symbol}, this method involves encoding evolutionary trajectory with continuous variables into infinite sequences of discrete symbols \cite{1995An,1988chaos}, providing a coarse-grained representation of chaotic attractors as well as other observable quantities. Through effective symbolic partitioning, one can derive the topological feature from these sequences that closely approximate the original dynamical structure. This process resembles recursive graph analysis\cite{1987Recurrence,MARWAN2007237}, thereby preserving crucial deterministic dynamics information essential for rigorous statistical analysis \cite{2017Elements,1985Ergodic}.
Additionally, there has been much work connecting the symbolic dynamics to the practical application, such as classification of unstable periodic orbits\cite{2009Complex,DONG20221}, entropy analysis\cite{KARAMANOS1999}, synchronization analysis\cite{2008Symbolic}, parameter reconstruction of chaotic systems\cite{X1995Symbol}, encrypted communication with chaos \cite{Hayes1993CommunicatingWC,1994Experimental} and gene network model\cite{2001Symbolic}. 

In the realm of symbolic dynamics, the exploration of partitioning the state space within complex chaotic regions is a widespread investigation field.
For noninvertible one-dimensional maps, it is demonstrated
that there is a rigorous way to locate symbolic partition boundaries, i.e. the position of all critical points \cite{2010chaos}.
However, a significant challenge in symbolic dynamics arises when attempting to partition the state space of chaotic maps in dimensions greater than one. Finding effective partitions that accurately capture the dynamics has proven difficult, limiting the method's applicability in higher-dimensional systems.
In the field of symbolic dynamics, despite the variety of partitioning methods, they generally fall into two main categories. One category involves analyzing the geometric structure of chaotic regions, while the other starts from symbolic sequences, focusing on the uniqueness of sequences at different positions to develop algorithms.
For first set of the methods, the study of geometric structures heavily relies on analyzing stable and unstable manifolds. 
It is conjectured that a generating partition passes through the primary homoclinic tangencies (PHT) between the stable and unstable manifold \cite{1985Generating,1989On}.
This method is not only implemented in dissipative systems but also in some conservative systems\cite{1995A}. Misha Chai \cite{2021Symbolic} proposed a new approach that focuses only on unstable manifolds of the map, which is based on the stretching and folding mechanism of chaotic generation\cite{2002chaos}, avoiding the analysis of a complicated tangles between stable and unstable manifold. This approach identifies iteration genealogy of folding points and find that the points coincide with positions of the PHT. However, the set of above methods are based on fully known deterministic dynamics, making them generally unsuitable for handling noisy stochastic systems. In addition, these are difficult in constructing the stable and unstable manifolds in high dimensional chaotic state space.
The other category of methods employs the geometric and topological structures without requiring exact knowledge of the dynamical equations, instead starting directly from experimental data, even if the data contains noise. 
Some of them rely on unstable periodic orbits (UPO) densely embedded in the chaotic attractor, to a generate unique symbol sequence \cite{1991Model,1994Progress}. Ruslan and Ying Cheng \cite{2000Estimating} proposed an efficient strategy for determining the generating partitions that coarse partition of chaotic attractors is typically revealed by a relatively small number of short UPO and refined partition is obtained by optimizing a set of proximity functions.
Topological analysis \cite{1994Combining,1999From,2000From} is also used to divide the state spaces of chaotic system into different areas,
in the light of the reasonable idea that topological invariants are compatible with symbolic names to ensure dynamic relevance.

There are other methods proposed based on only a time series, obtaining a refined empirical partition even if the noise exists. Althougth it is also widely recognized that generating partitions do not exist in the presence of additional noise for experimental chaotic time series datasets \cite{1982Symbolic,1983Symbolic}, it is rational to partition corresponding deterministic chaotic systems into different symbolic regions
based on these noisy disturbed data. Kennel and Buhl \cite{2003Estimating,2005Statistically} proposed a partition scheme from a time series which applies minimization the number of “symbolic false nearest neighbors” to fix and refine the region specified by a finite block of symbols via stochastic optimization techniques. Meanwhile, symbolic shadowing algorithm was proposed \cite{2004Estimating,2013Estimating} and optimized\cite{2018Empirical,2020The}, which obtains several good candidates for minimal generating partitions. 
These approaches overcome the drawbacks of the first category while also expanding the application scope of symbolic dynamics. However, they still have many shortcomings: Firstly, due to issues with computational complexity and workload, it is often difficult to find the UPOs\cite{1997Finding,1997Detecting} with sufficient accuracy and quantity to apply the methods. In addition, the high dimensional complex topology is difficult to visualize. Finally, the optimum partition is hardly selected from various candidate partitions under certain conditions.
Further exploration reveals that the spectrum of Koopman operator provides a heuristic link to the symbolic dynamics.
Koopman operator\cite{1931Hamiltonian}, in the field of the operator theory, which describes evolution of functions defined in the state space of a dynamical system.
In recent years, the numerical approximations of the Koopman operator, such as the dynamical mode decomposition (DMD) \cite{2010Dynamic} and EDMD \cite{Williams2014ADA}, is widely used in various dynamical systems.
It have been extensively applied on different occasions including the fluid system \cite{2009Spectral,Bagheri2013,2011Applications}, building energy efficiency \cite{Eisenhower2010DECOMPOSINGBS,2012CREATING}, power systems\cite{2011Nonlinear,2012Nonlinear} and neural networks\cite{Mezi2020KoopmanOG}.
According to the above conjecture, Cong Zhang\cite{2022Phase} carried out symbolic partitions for chaotic maps based on properly constructed eigenfunctions of a finite-dimensional approximation matrix of Koopman operator. This exploration reveals a connection between the symbolic partition and the eigenfunctions of the Koopman operators.
Nevertheless, the approach is crude and primitive in that it has a limited applicability and can only be applied to a subset of relatively simple chaotic systems and the precision accuracy of critical point positions is not high in 2-D map even in the absence of noise.
The Koopman eigenfunctions of this method only provide a global identification, which is a main reason for this issue. The extreme points of the obtained eigenfunctions are quite a lot, some of which coincide with the pre-image and after-image points instead of the boundary points themselves. Due to hardly identifying boundary points from these, this method is suitable for symbolic partition some simple 1-D chaotic system but fails to precisely partition the state space of multidimensional system which has complex local structures and multiple folding points.

In this paper, we propose a novel approach, combineing Koopman with stretching and folding mechanism of chaotic generation, to construct the symbolic boundary of chaotic regions. We construct a finite-dimensional approximate matrix of Koopman operator and calculate Koopman left eigenfunctions and focus on the functions whose eigenvalue are approximately equal to zero and positive-negative oscillations occur internally. Then, we can localize the whole region into the oscillation subregion and take it as the coarse symbolic boundary subregion. Then we refer to the function as Valid Left Eigenfunctions with Zero, abbreviated as VLEZ.
We can further the classify VLEZ based on the number of oscillations.
Based on this phenomenon, we propose Koopman Analysis (KA) method to obtain the VLEZ and then take symbolic partition for chaotic map.
In addition, we can refine the coarse symbolic boundary via combination affine transformation and  KA method on the coarse symbolic boundary. We refer to the modified method as Generalized Koopman Analysis (GKA).
For complex cases, primarily involving multimodal univariate chaotic map and multivariate chaotic map, we cannot obtain the VLEZ at low resolutions which means that primary coarse symbolic boundary still do not be obtain. Fortunately, some left functions similar to VLEZ except that the eigenvalues are unequal to zero and just close to zero. We refer to them as characteristic eigenfunctions (CLE). We consider the oscillation of the CLE as the estimated symbolic boundary. Then,  we redistribute basis functions based on the estimated symbolic boundary.
Under redistributed basis functions, we take KA method and obtain valid primary coarse symbolic boundary for the multimodal and multivariate chaotic maps, even in the presence of observational noise. Likewise, we can also refine the coarse symbolic boundary via GKA method.

The rest of the paper is divided into the following sections:
Section II firstly introduces the concept of Koopman operator and proposes its numerical representation, together with the definition of left eigenfunctions and the method for obtaining them.
Then, we discuss the mechanism of symbolic partition based on unimodal chaotic maps and take KA method on the chaotic series. Then, we obtain the coarse symbolic boundary based on one type of the Koopman left functions, referred to as VLEZ. Finally, we refine the coarse boundary subregion  using the GKA method, which integrates the KA method with affine transformation. 
In Section III, we apply proposed KA method to obtain CLE and VLEZ and further take GKA method to refine the coarse boundary for addressing the symbolic partition of the complex chaotic map. As a result, the precise boundary is obtainable for various complex chaotic series, including the multimodal map, Hénon map, Duffing oscillator and a three-dimensional hyperchaotic map proposed by Baier and Klein\cite{generalized}.
Naturally, the validity, applicability and robustness of the proposed method are all confirmed.
The paper is summarized in the Section V where we evaluate the proposed method and compare it with previous methods.

\section{Methodology }
	\label{sec:koopman}
In this section, we propose a new scheme for symbolic partition based on a generalization of  Koopman operator.  
Firstly, we introduce the spectral decomposition of the Koopman operator and its
numerical representation in section \ref{sec:koopman_rep}. 
Then, we explain the underlying principle and the actual procedure for symbolic partition based on the Koopman operator theory. 

\subsection{ Left Egienfuction of Koopman operator}
\label{sec:koopman_rep}

We consider a chaotic dynamical system in discrete time evolving on a manifold $\mathbb{R}^d$: for every state $\mathbf{x}\in \mathbb{R}^d$ 

\begin{equation}
	\mathbf{x_{\text{p}}}=T(\mathbf{x})
	\label{eq:evolution_define}
\end{equation}
where $T(\cdot) :\mathbb{R}^d \rightarrow \mathbb{R}^d$ is a nonlinear map defined on $\mathbb{R}^d$. 
The Koopman operator is a linear operator that describes the evolution of the observable, which is a scalar-valued function.
\begin{equation}
	Uf(\mathbf{x})=f(\mathbf{x_{\text{p}}})
	\label{eq:Koopman_define}
\end{equation}
where $f(\cdot):\mathbb{R}^d\rightarrow \mathbb{R}$ is the observable function. 
Crucially, the $f(T(\cdot))$ is regarded as the new function of the state variable $\mathbf{x}$.
\begin{equation}
	Uf(\mathbf{x})=f_{p}(\mathbf{x})
	\label{eq:Koopman_define_2}
\end{equation}

The Perron-Frobenius operator $\mathcal{P}$ as the adjoint operator of the Koopman operator, is also a linear evolution operator. This operator describes the evolution of density distribution of the state points 

\begin{equation}
	\int_{A} \mathcal{P}\rho(\mathbf{x})d\mathbf{x} =\int_{T^{-1}(A)} \rho(\mathbf{x})d\mathbf{x}
	\label{eq:PF_define}
\end{equation}
where $A$ is the set of the state point while the $T^{-1}(A)$ is the pre-image of $A$.

Firstly, in order to obtain the left eigenfunctions, we need to construct the approximate matrix $U$ of the Koopman operator.
As long as we have a complete set of observables $f(\mathbf{x})$ and evolved functions $f(\mathbf{x_{\text{p}}})$, we can construct a well-defined Koopman operator. We only need to find a complete set of observable functions within this region.
We refer to the observable functions used for construction of Koopman operator as basis functions. The set of basis functions is approximate as a column vector set 

\begin{equation}
	K=
	\begin{bmatrix}
		{g_{1}(\mathbf{x_{\text{1}}})}&{g_{2}(\mathbf{x_{\text{1}}})}& \cdots &{g_{M}(\mathbf{x_{\text{1}}})}\\
		{g_{1}(\mathbf{x_{\text{2}}})}&{g_{2}(\mathbf{x_{\text{2}}})}& \cdots &{g_{M}(\mathbf{x_{\text{2}}})}\\
		\vdots & \vdots & \ddots & \vdots \\
		{g_{1}(\mathbf{x_{\text{n}}})}&{g_{2}(\mathbf{x_{\text{n}}})}& \cdots &{g_{M}(\mathbf{x_{\text{n}}})}\\
	\end{bmatrix}
	\label{eq_K}
\end{equation}
where $\mathbf{x_{\text{k}}}$, $\text{k}=1,2,\dots, n$, is the state points and the $g_{i}(\mathbf{x})$, $i=1,2,\dots, M$, is the basis function.

Similarly, we obtain a set of evolved observable functions $g_{i}(\mathbf{x_{\text{p}}})$ via evolving the state $\mathbf{x_{\text{k}}}$ to $\mathbf{x_{p\text{k}}}$ governed by Eq.\eqref{eq:evolution_define} represented as $L$

\begin{equation}
	L=
	\begin{bmatrix}
		{g_{1}(\mathbf{x_{p\text{1}}})}&{g_{2}(\mathbf{x_{p\text{1}}})}& \cdots &{g_{M}(\mathbf{x_{p\text{1}}})}\\
		{g_{1}(\mathbf{x_{p\text{2}}})}&{g_{2}(\mathbf{x_{p\text{2}}})}& \cdots &{g_{M}(\mathbf{x_{p\text{2}}})}\\
		\vdots & \vdots & \ddots & \vdots \\
		{g_{1}(\mathbf{x_{p\text{n}}})}&{g_{2}((\mathbf{x_{p\text{n}}})}& \cdots &{g_{M}(\mathbf{x_{p\text{n}}})}\\
	\end{bmatrix}
	\label{eq_L}
\end{equation}
where every observable function $g(\mathbf{x_{\text{p}}})$ is also represented as the $\tilde{g}(\mathbf{x})$ based on Eq.\eqref{eq:Koopman_define} and Eq.\eqref{eq:Koopman_define_2}.
Obtaining the $K$ and $L$, the approximative matrix of Koopman operator can also be defined as
\begin{equation}
	K\tilde{U}=L
	\label{eq:koop_U2}
\end{equation}
where $\tilde{U}$ is the approximative matrix. 
For the steady-state chaotic discrete system, the finite-dimensional matrix $\tilde{U}$ can be represented as the approximation of Koopman operator because the subspace of infinite-dimensional function space must be found if the evolution is in a finite region $\mathbb{R}^d$. 
Every column of $\tilde{U}$ is the coefficient vector of basis function set $K$ for obtaining one $\tilde{g}(\mathbf{x})$. 
Every row of $\tilde{U}$, also the column of $\tilde{U}^{T}$, is the coefficient vector of basis function set $K$ for obtaining the density evolution function based on Perron-Frobenius operator. For obtaining the special density evolution function,
we take spectral decomposition of $\tilde{U}$ 
\begin{equation}
	u\tilde{U}=\lambda u
	\label{eq:Koopman_eigen}
\end{equation}
where $u$ is left eigenvector. We obtain the special coefficient column vectors $u$ for obtaining the left eigenfunction of $U$ represented as $Ku$, which is a numerical approximation of special density distribution $\rho(\mathbf{x})$ resulting from that $\mathcal{P}$ is the adjoint operator corresponding to $U$.

\begin{equation}
	Ku\tilde{U}= \lambda Ku
	\label{eq:Koopman_F}
\end{equation}

\subsection{ Mechanism of symbolic partition} \label{mech}
\subsubsection{Symbolic dynamics}

\begin{figure}[!htbp]
		\centering
		\includegraphics[width=8.5cm]{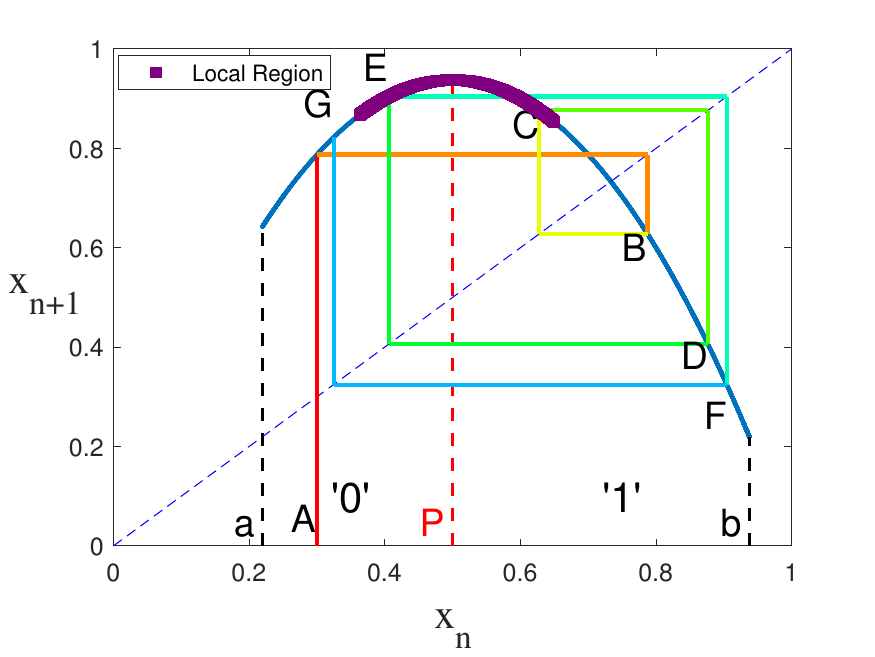}
		\caption{The graph relating $\mathbf{x}$ to $\mathbf{x_\text{{p}}}$ of chaotic unimodal map Eq.\eqref{eq:one_logistic_eq} with $\alpha$=3.75 
		}
		\label{fig:symbol_origin}
	\end{figure}

To better describe the symbolic dynamics and illustrate the mechanism of symbolic partition in chaotic systems, we should take an ideal chaotic map as an example. Here, we chose logistic map
\begin{equation}
	x_{n+1}=\alpha x_{n}(1-x_{n})
	\label{eq:one_logistic_eq}
\end{equation}
where $\alpha$ is a parameter that controls the folding. When $\alpha$ takes on certain specific values, the map becomes chaotic and the attractor changes from a finite to an infinite set of points. We take $\alpha$=3.75 as a parameter of chaotic map as is drawn in Fig.\ref{fig:symbol_origin}(a). $P$ is the only extreme point of the map, which is obviously the folding point. 
From Fig.\ref{fig:symbol_origin}(a), we find that the overlap of the two regions after folding make the points from different regions difficult to identify. 
In order to avoid the problem, we should give the initial symbolic partition, the two divided segments of the whole region $\mathbb{R}^d$ are referred as the different symbolic regions, $'0'$ and $'1'$, separately.
Assigning the symbol into initial point $A$ and the evolved points $BCDEFG$, we might obtain the symbolic sequence as $'0111010'$, which is a unique symbolic sequence on a smaller interval. If we increase the number of the symbolic sequence, the unique symbolic sequence could be assigned into every state point. 
Therefore, if we find the accurate symbolic boundary, the symbolic dynamics of the chaotic series could be constructed, greatly simplifying the representation of the original trajectory and thus its analysis.
Based on the above analysis, the pre-image of the folding point is the boundary of the symbolic partition, so we also refer the point as the boundary point. 

	Understanding the boundary point for symbolic partition is the pre-image of the folding based on the above analysis,
	we can apply the Koopman (KA) method for identification of the folding point for symbolic partition.

	\subsubsection{Symbolic partition with Koopman operator}
\begin{figure}
		
		\centering
		\includegraphics[width=8.5cm]{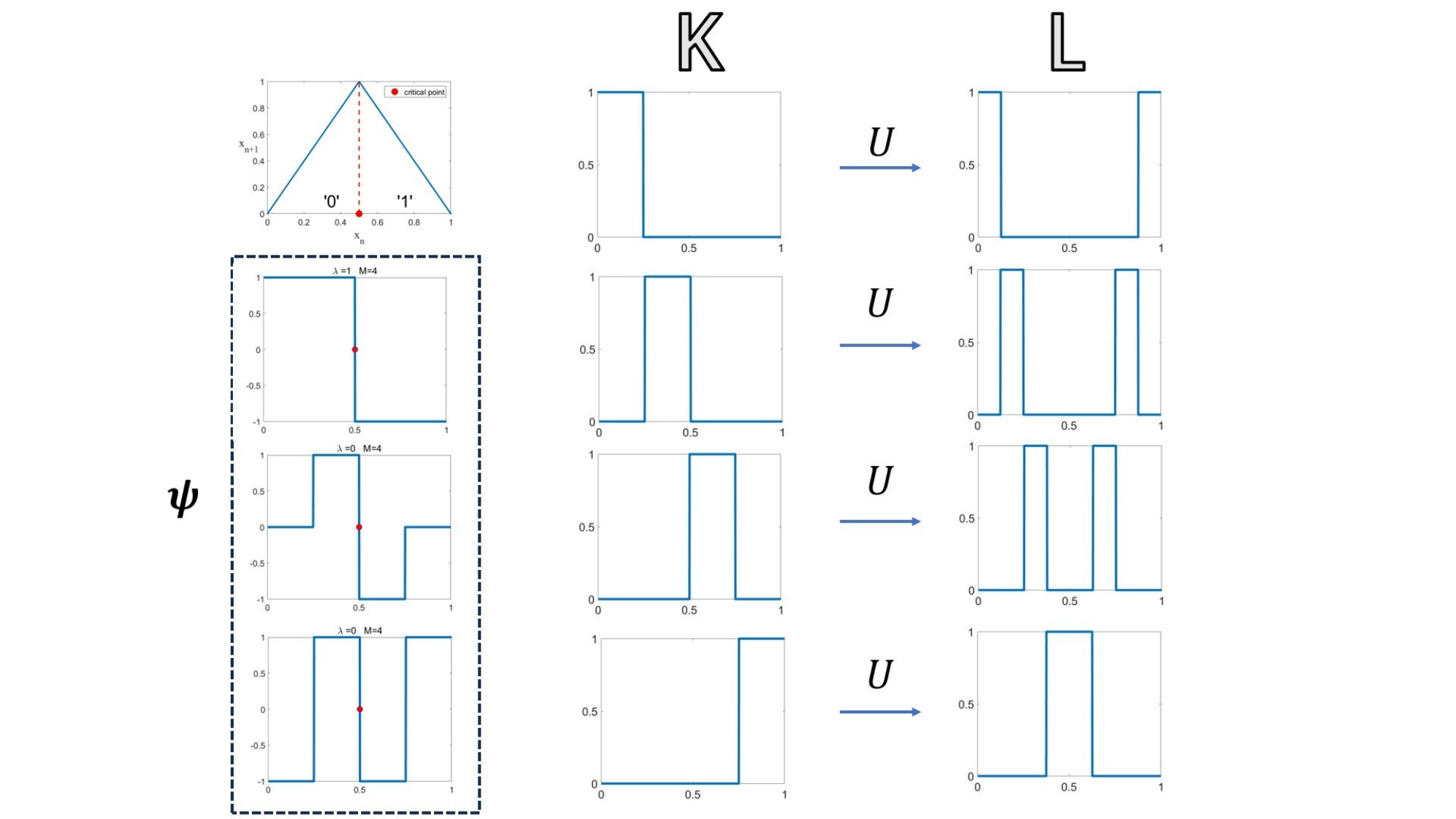}
		\caption{The four basis functions $K$ and evolved functions $L$
			and three typical left eignfunctions of the Koopman operator $U$ from the tent map Eq.\eqref{eq:tent}. 
		}
		\label{fig:mechanism}
	\end{figure}

	For simply explaining mechanism of symbolic partition via Koopman analysis, we take the tent map 
	\begin{equation}
		x_{n+1}=1-2\left|x_n-\frac{1}{2}\right|=
		\begin{cases}
			\begin{aligned}
				2x_n&,\ x\in [0,\frac{1}{2})\\
				2-2x_n&,\ x\in [\frac{1}{2},1]
			\end{aligned}
		\end{cases}
		\label{eq:tent}
	\end{equation}
	as the ideal unimodal map. As a one-dimensional piecewise linear map shown in Fig.\ref{fig:symbol_origin}(b), tent map Eq.\eqref{eq:tent} is still similar to the logistic map in that they are both unimodal and therefore have only one folding point during one-step evolution.
	Since the folding is complete after one-step evolution in such case, there is the primary mechanism of symbolic partition via Koopman analysis. First of all, we choose a group of simple observable functions as the basis functions, that is the rectangular function Eq.\eqref{eq_rect}.
	The basis functions $K$ and its evolved functions $L$ in this case are shown in Fig.\ref{fig:mechanism} for $M=4$. 
	we obtain the transformation matrix 
	
	\begin{equation}
		\tilde{U}=\begin{pmatrix}
			0.5 & 0.5 & 0 & 0  \\
			0 & 0 & 0.5 & 0.5  \\
			0 & 0 & 0.5 & 0.5  \\
			0.5 & 0.5 & 0 & 0  \\
		\end{pmatrix}\label{eq:Koopman_U4}
	\end{equation}
	based on Eq.\eqref{eq:koop_U2}.

	Then we take spectral decomposition of $\tilde{U}$ based on Eq.\eqref{eq:Koopman_eigen} and obtain left eigenvectors as the coefficient with only two eigenvalues $\lambda$ namely $1$ and $0$. The eigenvector with $\lambda=1$ is $[1,1,1,1]$ while eigenvectors with $\lambda=0$ are degenerate. There are three typical eigenvectors with $\lambda=0$ after analytic calculation, namely $[1,1-1,-1]$, $[0,1,-1,0]$ and $[-1,1,-1,1]$, which are all displayed in Fig.\ref{fig:mechanism}. 
	We refer to the Koopman left eigenfunction with $\lambda=0$ as Left Eigenfunction with Zero (LEZ).

	For the first one, the subregion whose function values are all positive in interval $[0,0.5]$, denoted as $R_{A}$, is the pre-image of the whole region $\mathbb{R}^d$. The other subregion whose function values are all negative in interval $[0.5,1]$, denoted as $R_{B}$. We refer to $R_{A}$ as positive-valued region while $R_{B}$ as negative-valued region. The positive-valued region and adjacent negative-valued region construct an oscillation. Thus, we also refer to the whole region as the oscillation subregion.
	According to Eq.\eqref{eq:PF_define}, the eigen-distribution $\rho(\mathbf{x})$=0 on $\mathbb{R}^d$  after one iteration for eigenvalue $\lambda=0$. 
	This phenomenon indicate that the two subregions can overlap completely after one step evolution. 
	We can come to a conclusion that there is a folding in this region based on the oscillation composed of one positive peak and one negative peak. However, we do not find the position of pre-image of folding, served as symbolic partition boundary in that the oscillation covers the whole region. We only identify that the symbolic partition boundary appears the oscillation subregion.
	For the second one, we can find that the oscillation lies on localized subregion, which means that we can obtain the coarse position of symbolic boundary. Meanwhile, the rest two subregions whose function values are all zero are both referred as the zero-valued subregions. Thus, we collectively refer to the positive-valued region and negative-value region as nonzero-valued region. We refer to the zero-valued subregion surrounding the oscillation subregion as non-oscillation subregion.
	The subregion indicates that no coarse symbolic boundary exists in this place. 
	Therefore, we can obtain the coarse boundary based on this type of LEZ.
	For the last one, we can find that there are multiple oscillations where the oscillation also covers whole region and no zero-valued subregion exists just as the first one.
	Based on analysis of above typical LEZ, we should obtain the LEZ whose oscillations appear in the inner subregion instead of whole region or edge surrounding subregion. Only localized subregion just as the second one are valid for obtaining coarse symbolic boundary subregion when addressing the symbolic partition. 
	Therefore, we propose the Valid Left Eigenfunction with Zero, also referred to as VLEZ.
	The VLEZ is the left eigenfunction of Koopman operator with $\lambda=0$ and there are oscillations  which only located in the inner subregion.
	In this article, considering numerical computation errors, the module of eigenvalue of VLEZ can be permitted approximately equal to zero with an error within $10^{-5}$ but ensure one of the three minimum eigenvalues. In addition, the function value of zero-valued region can be permitted in scope of $1/10$ of the maximum and minimum of oscillation subregion.
	If the coordinate range of a tiny zero-valued subregion is less than one-fifth of that of the oscillation region, it may be allowed to appear within the oscillation subregion, for example, as the tiny zero-valued subregion shown in Fig.\ref{fig:symbol one}(b), where the side lengths of the bounding rectangles of each coordinate are all less than one-fifth.

	Here, we further classify the VLEZ based on the type of the oscillation region 
	as unitary VLEZ (UVLEZ) and multiple VLEZ (MVLEZ). The UVLEZ encompasses the valid unitary oscillation region, while the MVLEZ possesses the valid multiple oscillation region.
	Specifically, valid unitary-oscillation region is the unitary-oscillation region where the ratio of the modulus of the minimum to the maximum of the whole regions should be limited in the scope $[\frac{1}{2},1]$.
	The multiple-oscillation region is set as the region composed of more than one unitary-oscillation subregion. 
	To further demonstrate the above conjecture and obtain the VLEZ effectively, we apply the proposed KA method to partition the series from the logistic map Eq.\eqref{eq:one_logistic_eq}.
	We take the proposed KA method to find the boundary point.
	To ensuring the validity of the basis functions and eigenfunctions, we should limit the bottom of $n$. In this article, we set the condition $n\geq100$.  In addition, we also should limit $M$ to ensure that the number of the basis function is quite smaller than that of state points. Here, we set the condition $M\leq5n$. The above constraints are adopted for all following cases.
	Specifically, we take the 10001 successive evolving points. The former $n=10000$ points are the $\mathbf{x}$ while the later $n$ points are the $\mathbf{x_{\text{p}}}$. We take rectangle basis functions with the parameters $n=10000$ and $M=5$. We obtain the UVLEZ as shown in Fig.\ref{fig:symbol one appd}(a) in Appendix A. We only ensure that the position of the symbolic boundary is in scope of the oscillating subregion. Therefore, we increase resolution via setting $M=50$ and then obtain the MVLEZ as is shown in Fig.\ref{fig:symbol one}(a). 
	
	\subsubsection{Symbolic partition with auxiliary mappings}
	
\begin{figure*}[!htbp] 
		\begin{minipage}[t]{0.48\linewidth}
			\centering
			\begin{subfigure}[t]{\textwidth}
				\centering
				\includegraphics[width=8.5cm]{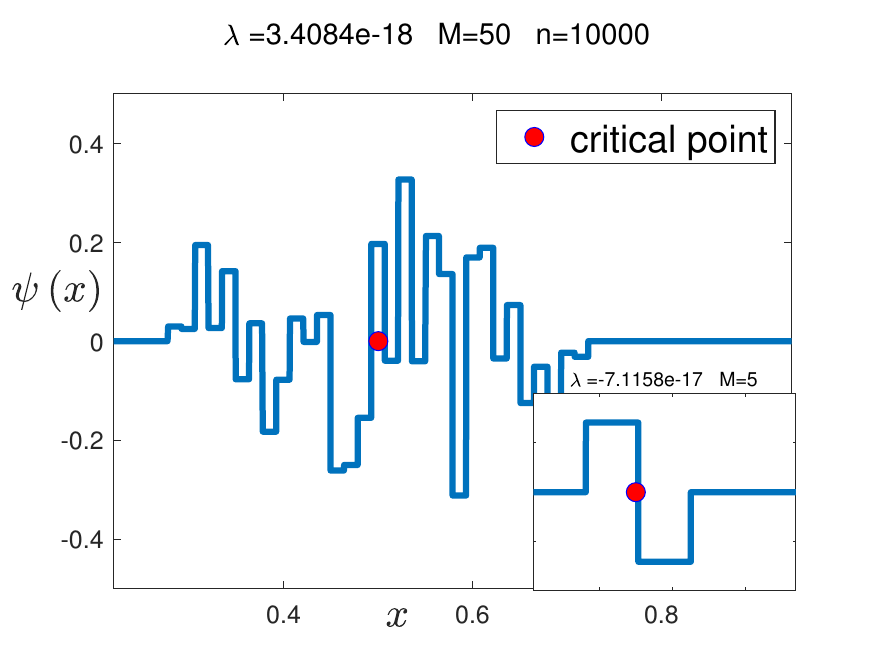}
				\caption{}
			\end{subfigure}
		\end{minipage}
		\hfill
		\begin{minipage}[t]{0.48\linewidth}
			\centering
			\begin{subfigure}[t]{\textwidth}
				\centering
				\includegraphics[width=8.5cm]{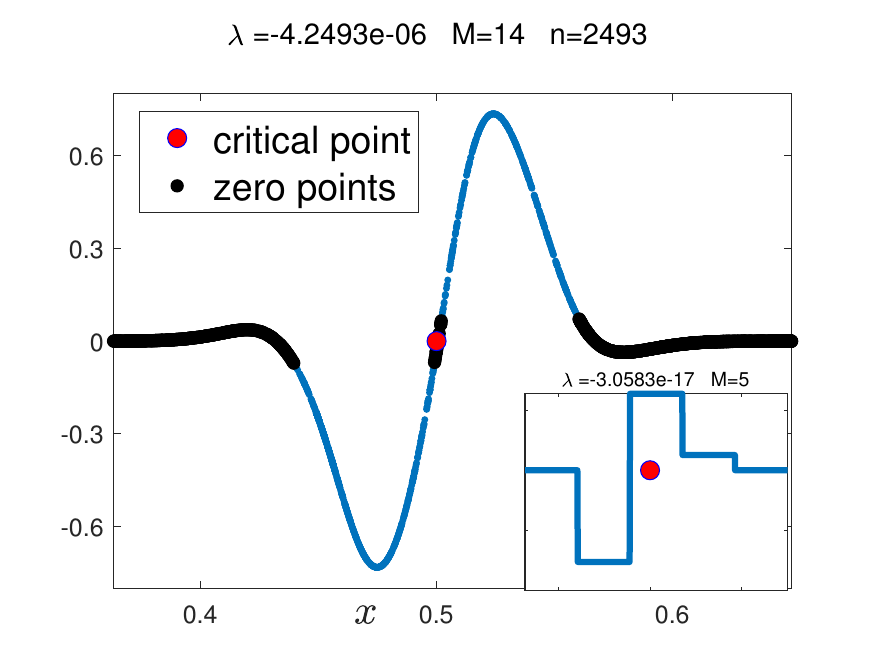}
				\caption{}
			\end{subfigure}
		\end{minipage}
		\caption{The KA and GKA methods for the chaotic unimodal map.		
			(a) One MVLEZ in the whole region via KA method.
			(b) The UVLEZ in the localized region via GKA method.
		}
		\label{fig:symbol one}
	\end{figure*}

	According to the two VLEZ, we find that it is difficult to obtain precise symbolic boundary as a result in still broad symbolic boundary subregion by increasing resolution $M$.
	we should find a new method for precise symbolic boundary.
	Since the folding points do not exist in zero-valued regions, we neglect two zero-valued subregions. 
	Then, we localize the oscillation region of the VLEZ based on $M=5$ shown as the purple region shown in Fig.\ref{fig:symbol one}(a).
	
	However, we observe that the point sets before and after evolution reside in different subspaces. In order for the Koopman analysis to apply, we need to move them to the same subspace without changing the nonlinearity, necessitating affine transformations. Taking the local purple region as an example, we refer to the region that contains the original state points $\mathbf{x}$ as $\mathbf{R}$, which is the interval $[0.3631,0.6505]$. The corresponding region of after one-step evolution $\mathbf{x_{\text{p}}}$, denoted as $\mathbf{R_{\text{p}}}$, which is the interval $[0.8532,0.9375]$. 
	We choose the target points of $\mathbf{x}$ and $\mathbf{x_\text{{p}}}$ during the translation in univariate series. The target points are always the left endpoints in univariate series and therefore are referred to as  $\mathbf{x_{\text{min}}}$ and $\mathbf{x_{\text{p\_min}}}$.
	
	We intend to choose only one region from $\mathbf{R_{\text{p}}}$ and $\mathbf{R}$ as the $\mathbb{R}^d$ and thus take pretreatment. Specifically, the transformation of $\mathbf{x_{\text{aff}}}$ which is linear and does not interfere with the folding process but keep $\mathbf{x_{\text{aff}}}$ and $\mathbf{x_{\text{p}}}$ in the same region $\mathbf{R_{\text{p}}}$. Therefore, we should take affine transformation to $\mathbf{x}$. The equation is
	\begin{equation}
		\mathbf{x_{\text{aff}}}=\frac{l(\mathbf{R_{\text{p}}})}{l(\mathbf{R})}(\mathbf{x}-
		\mathbf{x_{\text{min}}})%
		+\mathbf{x_{\text{p\_min}}}
		\label{eq:affine}
	\end{equation}
	where $\mathbf{x_{\text{aff}}}$ is the aim point after affine translation. $\frac{l(\mathbf{R_{\text{p}}})}{l(\mathbf{R})}$
	is the scaling term. It is worth noting that the affine transformation Eq.\eqref{eq:affine} is just applied to univariate series. For multivariate series,  $\mathbf{R_{\text{p}}}$ and $\mathbf{R}$ are always not linear region, which causes the invalidity of Eq.\eqref{eq:affine}. The affine transformation for  multivariate series are explained in detail in Appendix B.

	Based on the above analysis, we simplify the form of the Eq.\eqref{eq:affine} to become the equation
	\begin{equation}
		\mathbf{x_{\text{aff}}}=S\mathbf{x}+L
		\label{eq:affine_abb}
	\end{equation}
	where the former term $S$ is the scaling term while the latter term $L$ is the term of the translation process corresponding to two terms of the Eq.\eqref{eq:affine}.

	Similar to the whole map in Fig.\ref{fig:symbol_origin}(a), the relation graph of the $\mathbf{x_{\text{aff}}}$ and $\mathbf{x_{\text{p}}}$ resembles a unimodal chaotic map as shown in Fig.\ref{fig:symbol one appd}(a) in Appendix B, making it feasible to perform KA in this case.
	Performing KA analysis, we obtain the UVLEZ based on $\mathbf{x_\text{{p}}}$. Then we can obtain the UVLEZ based on the original localized region $\mathbf{R}$ via reversing the variable $\mathbf{x_{\text{aff}}}$ into $\mathbf{x}$ based on the reverse-transformation of the Eq.\eqref{eq:affine} in subplot of Fig.\ref{fig:symbol one}(b).
	
	Due to original points $\mathbf{x}$ and the evolved points $\mathbf{x_\text{{p}}}$ are in different regions, this method differs from the Koopman Analysis to some extent. Thus, the symbolic partition method is referred to as Generalized Koopman Analysis, abbreviated as GKA.
	To illustrate that our VLEZ is independent of the specific basis function type, we replace the rectangle functions with Gaussian functions Eq.\eqref{gauss}. Specifically, we chose $M-2$ equidistant points and 2 end points to divide up $\mathbf{R_{\text{p}}}$  into $M$ segments and set the $M$ points as grid points. We set the length of the segment as the wave packet radius. The state points are $\mathbf{x_{\text{aff}}}$ and $\mathbf{x_{\text{p}}}$ whose number are both $n=2493$. Initially, we set $M=3$ and thus $r_{w}=0.0283$ based on $l(\mathbf{R_{\text{p}}})=0.0849$ and VLEZ can be obtained. Then, we increase $M$ one by one and we eventually obtain VLEZ when $M=14$ and $r_{w}=0.0061$ as shown in Fig.\ref{fig:symbol one}(b). 
	Unlike the oscillation of UVLEZ based on rectangle functions, that of UVLEZ based on Gaussian functions is so smooth that we can find the tiny zero-valued region inside oscillation. We can take this zero-valued region as the symbolic boundary instead of whole oscillation region in that the two nonzero-valued subregions are the both side of pre-image folding while the symbolic boundary point is not folding. Therefore, it is reasonable to take the tiny zero-valued oscillation region as the symbolic boundary points.
	We also refer to the points of the tiny zero-valued oscillation region as zero-crossing points.

	Here, we find that the number of Gaussian functions must be sufficiently larger than rectangle functions when the VLEZ appears. However, we can obtain more precise symbolic boundary via Gaussian functions.   
	Therefore, for whole regions, we need few basis to simply find VLEZ and obtain every coarse boundary. In face of refining the coarse boundaries, we might choose Gaussian functions $g_{GA}$ as basis functions to simplify refinement process and obtain more precise position. 
	
	Considering the efficiency, we should increase $M$ one added by one when using rectangle functions while increase $M$ ten added by ten when using Gaussian functions.
	When we take GKA method on localized non-chaotic region, we can obtain another LEZ.
	We perform an affine transformation locally, which is equivalent to constructing a steady-state region in one step or multiple steps. 
	Unlike addressing the whole chaotic regions, this steady-state region cannot only be the chaotic map in Fig.\ref{fig:symbol one appd}(a) but also an ordinary attractor in Fig.\ref{fig:symbol one appd}(b) where the original localized region is [0.3631,0.5000].
	We take GKA on the region and obtain a LEZ as the subplot in Fig.\ref{fig:symbol one appd}(b). We find that the oscillation subregion of VLEZ are the surrounding of attractor $x=0.9375$ while the ordinary zero-valued region are the residual region. 
	The LEZ is invalid in that there are no folding exists after one step evolution. Because the LEZ consists of one oscillation subregion and non-oscillation subregions, the LEZ is the interference when we take GKA method. 
	To avoid the interference of the invalid LEZ when taking GKA method, we need to ensure that the selected localized region must contain one folding after one-step evolution before conducting the GKA method. 
	However, the oscillation subregion of invalid LEZ is the surrounding of edge and indicate that the localized regions is one side of the symbolic boundary.
	Therefore, if we obtain the invalid LEZ, we can enlarge the former localized region and can obtain the localized chaotic map and VLEZ.

	The above is the primary mechanism of coarse symbolic partition via KA method and refinement of the symbolic boundary via GKA method for partitioning the one-dimensional chaotic map.
	The mechanism is supposed to be generalized to the complex and multivariate chaotic series. In the following sections, we use the KA and GKA method for symbolic partition of the complex multivariate series.


	\section{Result and Discussion}
	\label{result00}
	
	In this section, we will apply the method to achieve symbolic partition in several typical examples in chaotic series originating from different nonlinear systems. The following examples are categorized into univariate and multivariate series. We firstly analyze the univariate series from chaotic multimodal map based on our proposed method and necesary pretreatment to obtain the valid symbolic boundary. Then we propose an corresponding explanation for the mechanism of the symbolic partition.
	Then several typical examples of complex multivariate series are partitioned by necesary pretreatment and the proposed method.

	\subsection{Analysis of the univariate chaotic series }
	\label{result_1}
	
	The partition starts with univariate series based on the folding is sole and complete so the result of symbolic partition is obtainable by the KA method just as the case in Section \ref{sec:koopman}. 
	After a successful symbolic partition of the simplest univariate chaotic series, we apply this method for complex univariate chaotic series which arise from one-dimensional multimodal map, which has multiple folding points. We construct the map 
	
	\begin{equation}
		x_{n+1}=x_{n}-5x_{n}^{2}+8x_{n}^{3}-4x_{n}^{4}
		\label{eq:two_logistic_eq}
	\end{equation}
	
	\begin{figure*}[!htbp] 		
		\begin{minipage}[t]{0.48\linewidth}
			\centering
			\begin{subfigure}[t]{\textwidth}
				\centering
				\includegraphics[width=8.5cm]{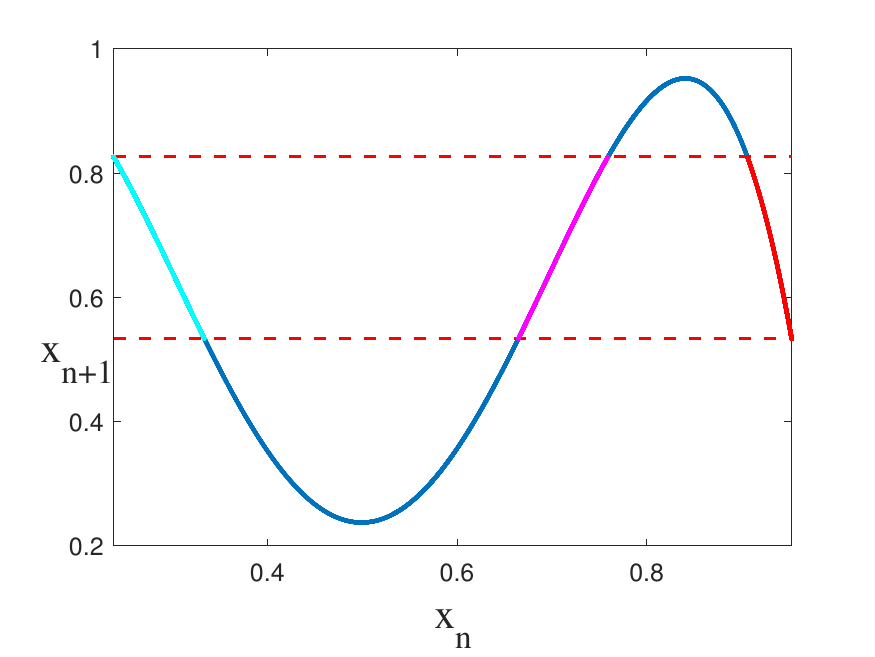}
				\caption{}
			\end{subfigure}
		\end{minipage}
		\hfill
		\begin{minipage}[t]{0.48\linewidth}
			\centering
			\begin{subfigure}[t]{\textwidth}
				\centering
				\includegraphics[width=8.5cm]{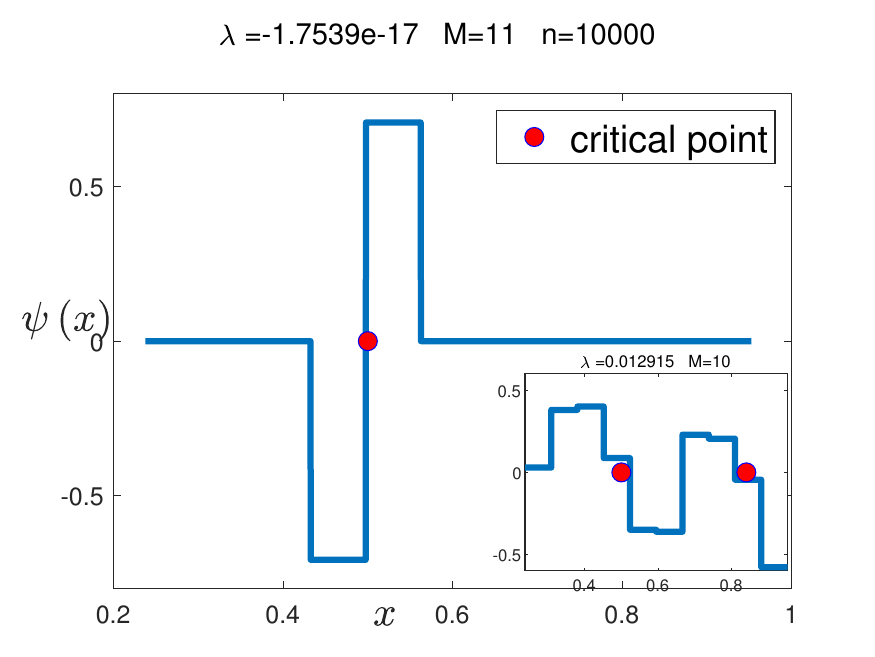}
				\caption{}
			\end{subfigure}
		\end{minipage}
		\hfill
		\begin{minipage}[t]{0.48\linewidth}
			\centering
			\begin{subfigure}[t]{\textwidth}
				\centering
				\includegraphics[width=8.5cm]{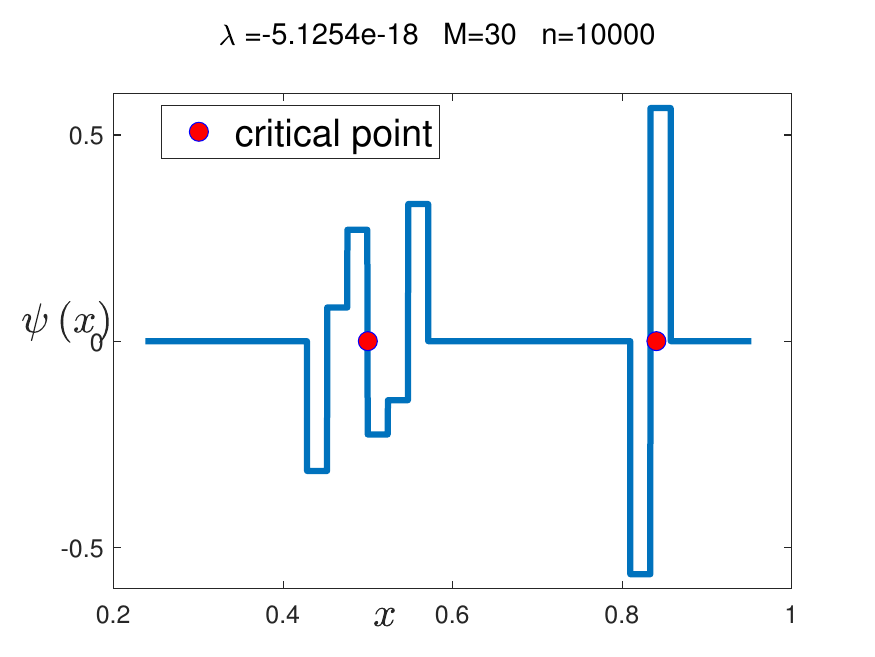}
				\caption{}
			\end{subfigure}
		\end{minipage}
		\hfill
		\begin{minipage}[t]{0.48\linewidth}
			\centering
			\begin{subfigure}[t]{\textwidth}
				\centering
				\includegraphics[width=8.5cm]{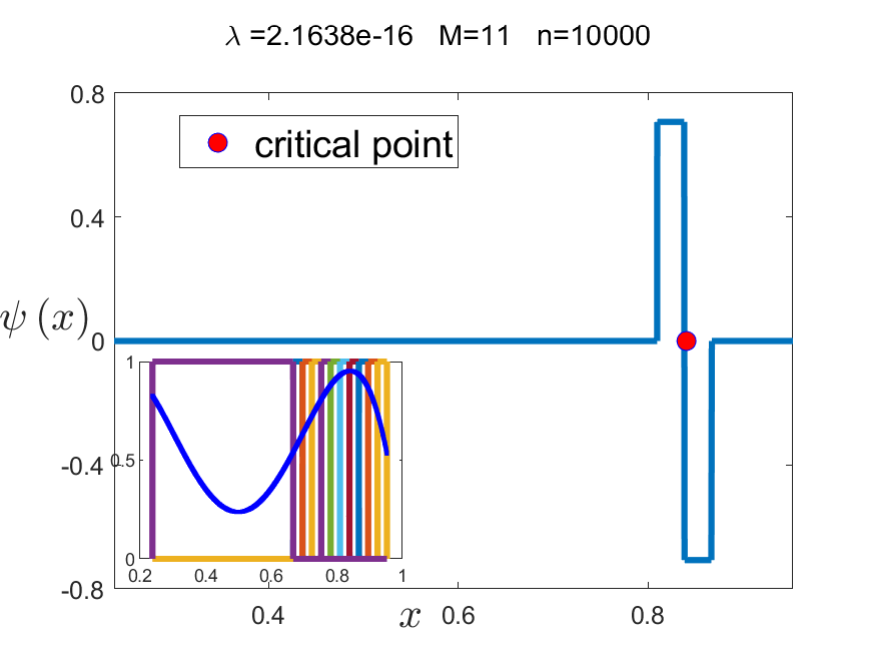}
				\caption{}
			\end{subfigure}
		\end{minipage}
		
		\caption{The basis functions of the chaotic map governed by the Eq.\eqref{eq:two_logistic_eq} and KA method for the chaotic multimodal map.
			(a) The graph relating $\mathbf{x}$ to $\mathbf{x_\text{{p}}}$ of chaotic map Eq.\eqref{eq:two_logistic_eq}
			(b) The UVLEZ to obtain the first symbolic boundary.
			(c)	The MVLEZ to obtain two symbolic boundaries.
			(d) The UVLEZ to obtain the second symbolic boundary.
		}
		\label{fig:symbol two}
	\end{figure*}
	
	The evolution function plot is shown in Fig.\ref{fig:symbol two}(a). The scope is $[0.2375,0.9526]$
	We find the map has two critical points. Due to the interference from other multiple foldings during the one-step evolution, we hardly obtain the VLEZ in this case under low resolution. We take KA method with the parameters $n=10000$ and $M$=10 and obtain one left eigenfunction with the minimal eigenvalue as the subplot is shown in Fig.\ref{fig:symbol two}(b), where the two coarse oscillations, which indicates that two folding exist. It is similar to the VLEZ except that eigenvalue is just close to zero. The module of the eigenvalue is in the scope of $[]10_{^5},0.1]$.
	We refer to the eigenfunction as the Characteristic Left Eigenfunction, abbreviated as CLE. The CLE is critical when we handle the multi-folding chaotic map, especially the multivariate chaotic map where the foldings are massive. We can obtain the oscillation of the CLE and refer to it as the estimated symbolic boundary. However, the estimated symbolic boundary are not considered as the primary boundary due to relatively large error of the eigenvalue of the CLE compared to VLEZ. Fortunately, we 
	can redistribute  the basis functions to ensure that the number of wave packets covering  estimated symbolic boundary subregion is large while number of that covering the non-symbolic subregions is small. Generally speaking, we set the number
	as 1 for every continuous non-symbolic subregion.
	
	Here, we find that the symbolic estimated boundary based on multiple oscillations of the CLE are almost are the whole region. Therefore,
	we just continue to increase the resolution and the UVLEZ are accordingly obtained when we set $M$=11 as shown in Fig.\ref{fig:symbol two}(b).
	However, we find the UVLEZ can find only one symbolic boundary. When we continue to enlarge $M$ to $30$, we can obtain the MVLEZ shown in Fig.\ref{fig:symbol two}(c) where we can find both two symbolic boundaries. However, we find the method based on increasing resolution is too complex to generalized to complicate multivariate chaotic map. Thus, we propose a novel method to find the other boundary.
	
	Consider the two coarse oscillation subregions of the  CLE and one coarse oscillation subregions of the former VLEZ, we should take KA method to obtain the other VLEZ as long as avoiding the interference from the former oscillation.
	Here, We set only one wave packet which cover the first boundary and surrounding zero-valued subregions [0.2375, 0.6666]. we chose $x=0.6666$ as the divided line based on the CLE as subplot shown in Fig.\ref{fig:symbol two}(b). Then, we take $M-1$ rectangle basis functions whose wave packets cover [0.6666,0.9526] as the subplot shown in Fig.\ref{fig:symbol two}(d). Then we obtain the UVLEZ just when $M=11$.
	Then, we take GKA method and refine two boundaries separately. Then, we should assign symbols to three subregions. We should take three symbols rather than two symbols in that we find three subregions overlap into one subregion after one-step evolution which are between the two red dashed lines. In case the third subregion does not exist, we can just assign two symbols even if two boundaries exist.
	Therefore, we should analysis the evolved subregion after obtaining the symbolic boundaries when we consider the number of symbols.		
	Overall, obtaining CLE and then redistribution wave packets of the rectangle functions are two central pretreatment steps. Then, we can generalized above pretreatment process to following complex multivariate chaotic series due to partial features similar to this case.
	We add small Gaussian white noise $\sigma$ whose mean is 0 and standard deviation is 1 into left terms of Eq.\eqref{eq:two_logistic_eq}.
	Under this parameter setting with the level of noise $\sigma$ = 0.005, the two boundary points can also be obtained which closely locate near the two boundary points in Fig. \ref{fig:symbol two}(b) and Fig. \ref{fig:symbol two}(d) indicating the robustness of the current technique.

	\subsection{Analysis of the multivariate chaotic series}
	\label{multi_example}
	
	In Section \ref{result_1}, the KA method and GKA method are both able to be applied to all one dimensional chaotic time series. In such condition, the folding is complete in one-step evolution. 
	
	For multi-dimensional chaotic series evolution, the folding is incomplete. As a result, the process continues after initial evolution step and the latter folding being more complete than the former. 
	Therefore, we must consider the multiple-step evolved subregion when we refine the coarse symbolic boundary based on one-step evolution KA method.
	Initially, we identify a coarse boundary region in that the folding in multi-dimensional chaotic region $\mathbb{R}^d$ is incomplete, where different subregions out of the boundary region converge into one outside subregion during the folding process. The coarse boundary region is called $\textit{primary boundary}$ and the process is called $\textit{primary folding}$.
	The primary folding is interfered with innumerable local foldings as a result that we just obtain CLE when we take KA method. Subsequently, we enhance the number of the wave packets covering the oscillation subregion and decrease the number of non-oscillation subregions to one. After completing above pretreatment process, we finally take KA method and accordingly obtain VLEZ.
	Here, we use the method in several examples of complex multivariate chaotic series.
	
	\subsubsection{series from H\'{e}non map I ($\alpha$=1.4, $\beta$=0.3)}
	\label{henonA}
	We take our method for symbolic partitioning a classic chaotic map, Hénon map, reported by Hénon \cite{M1976A}.
	The equation of the Hénon map is

	\begin{equation}
		\begin{split}
			x_{n+1} &=-\alpha x_{n}^{2}+y_{n}+1, 	\\
			y_{n+1} &=\beta x_{n}
		\end{split}
		\label{eq:henon1_eq}
	\end{equation}
	where $\alpha$ is a parameter that controls the folding and $\beta$ for the dissipation. With the conventional value $\alpha=1.4$ and $\beta=0.3$, the folding is relatively complete. Therefore, we take the chaotic series from  H\'{e}non map I ($\alpha=1.4,\beta=0.3$) as the first example of multivariate chaotic series. The result of symbolic partition is reported firstly by Grassberger and Kantz\cite{1985Generating}. We intend to analyze this case to verify the validity of our methods.

\begin{figure*}[!htbp] 
		
		\begin{minipage}[t]{0.48\linewidth}
			\centering
			\begin{subfigure}[t]{\textwidth}
				\centering
				\includegraphics[width=8.5cm]{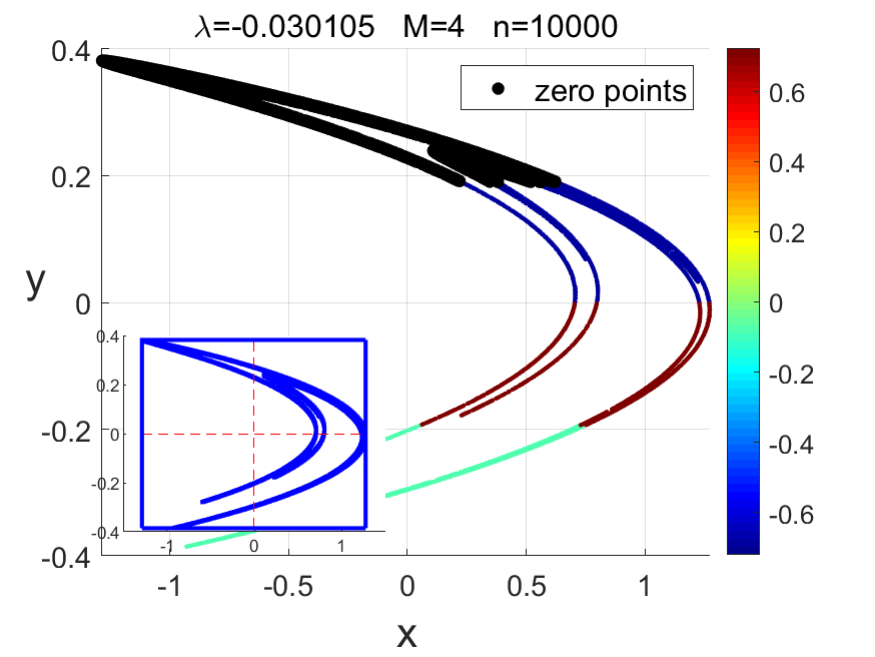}
				\caption{}
			\end{subfigure}
		\end{minipage}
		\hfill
		\begin{minipage}[t]{0.48\linewidth}
			\centering
			\begin{subfigure}[t]{\textwidth}
				\centering
				\includegraphics[width=8.5cm]{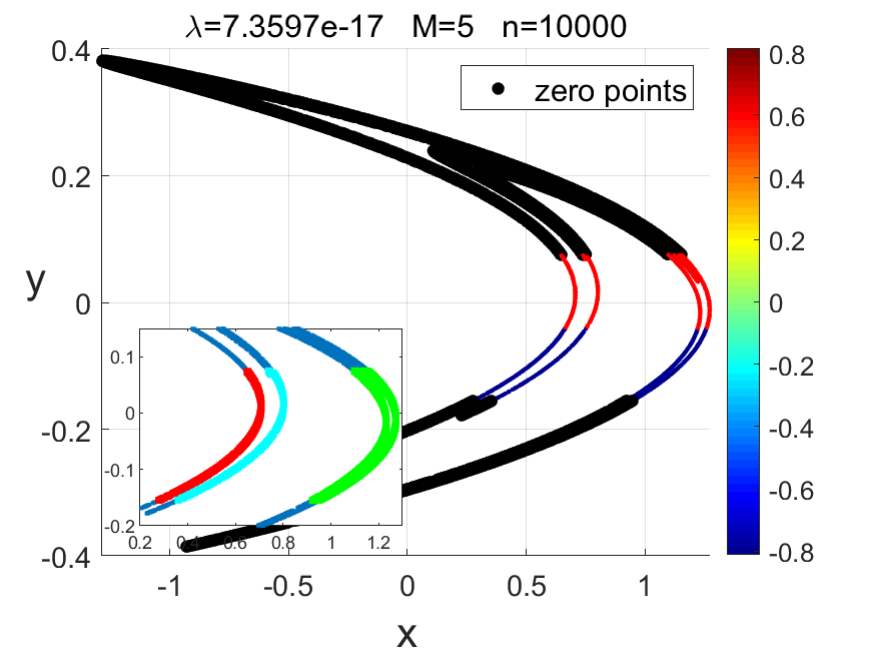}
				\caption{}
			\end{subfigure}
		\end{minipage}
		\hfill
		\begin{minipage}[t]{0.48\linewidth}
			\centering
			\begin{subfigure}[t]{\textwidth}
				\centering
				\includegraphics[width=8.5cm]{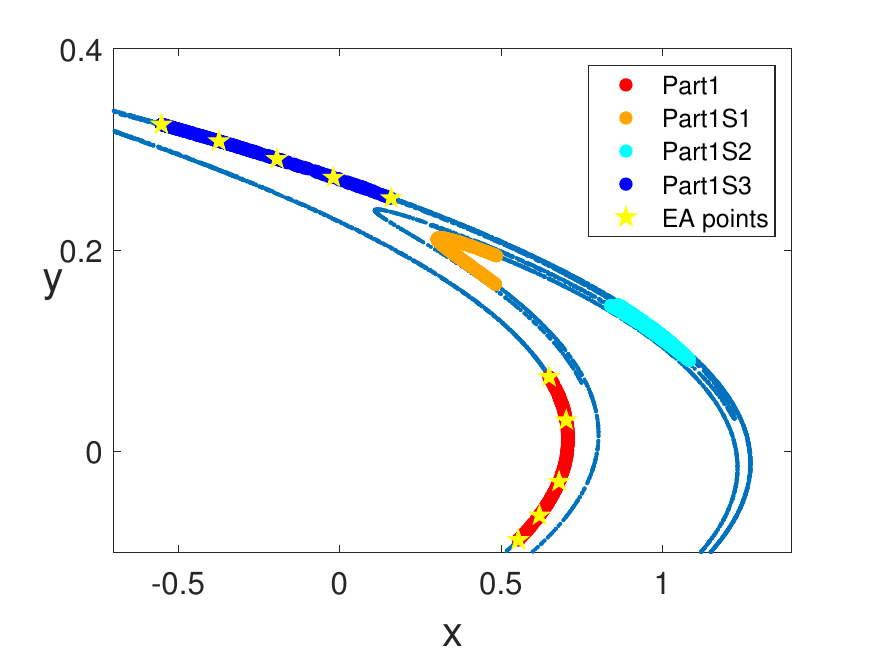}
				\caption{}
			\end{subfigure}
		\end{minipage}
		\hfill
		\begin{minipage}[t]{0.48\linewidth}
			\centering
			\begin{subfigure}[t]{\textwidth}
				\centering
				\includegraphics[width=8.5cm]{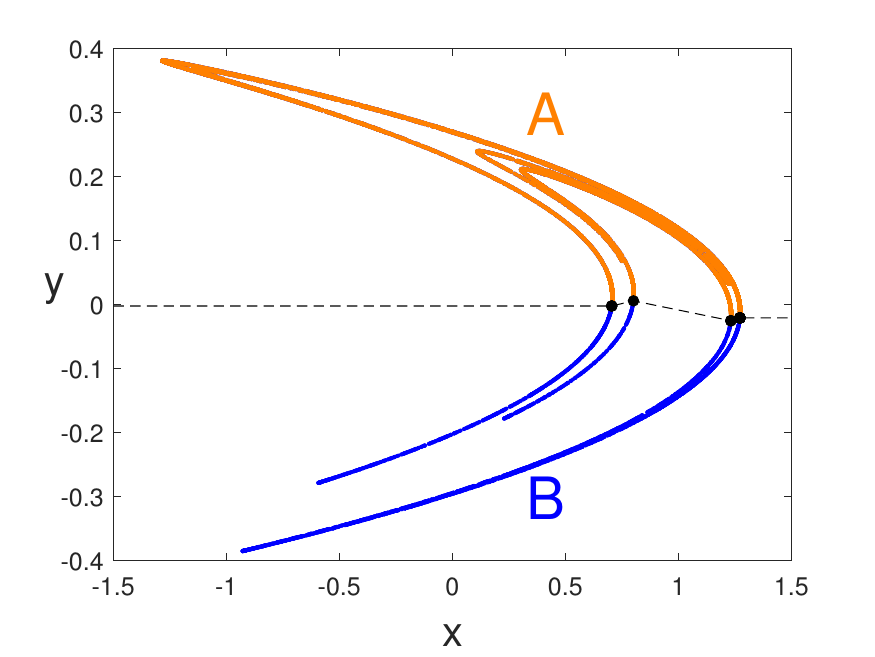}
				\caption{}
			\end{subfigure}
		\end{minipage}
		
		\caption{The KA method on the series governed by the H\'{e}non I map 
			(a)The CLE for primary symbolic boundary 
			(b)The VLEZ for primary symbolic boundary 
			(c)The multiple-step evolution of Part1
			(d)The eventually symbolic boundary and symbolic subregions
		}
		\label{fig:henon1_whole}
	\end{figure*}

	In multi-dimensional chaotic series, the folding is incomplete during one-step evolution and its complete folding need multi-step evolution. Therefore, the forward and backward regions of one folding region can interfere itself. In addition, the local narrow foldings are also obstacles for obtaining the symbolic boundary.
	As is represented in Fig.\ref{fig:henon1_whole}(a), the whole region is complex, comprising a few primary folding and innumerable different levels of local foldings. 
	Despite above problems, the region globally is continuous on a large scale so that we can find the coarse symbolic boundary.  
	Here, we generalize the KA method to the multivariate chaotic. Similarly to the one-dimensional chaotic map, we should construct the appropriate $K$ and $L$ for obtaining the VLEZ. 
	Specifically, the basis functions are the generalized bivariate rectangle functions.
	We construct a rectangular region whose boundary is tangent with the attractor. 
	
	Then, we chose both two coordinates (x and y direction) as the segmentation directions. The two numbers of the segments of coordinate directions $x$ and $y$ are $Mx$ and $My$.
	Initially, we should ensure $Mx=My$. Then, we set the parameters are from $Mx=My=2$ gradually increasing $Mx=My$. Then, when $Mx=My=2$, we obtain the $2\times2$ two-variate generalized rectangles as subplot shown Fig.\ref{fig:henon1_whole}(a). At that parameter, we can find that the continuity of the fractal region cannot be broken when $Mx$ are reset to 1 where we also ensure the regularity of every subregion. When $Mx=My\geq2$, $Mx$ should be reset to 1 based on the same cause.
	Therefore, we chose $M=My$ and gradually increase $M$ from 2.			
	As a result, we obtain the CLE when $M=4$ as is shown in Fig.\ref{fig:henon1_whole}(a). We can find the zero-valued subregion and consider it as the non-symbolic subregion and take rest subregion as the symbolic subregion. Based on above phenomenon, we can redistribute the wave packets of the generalized rectangle functions in an analogous manner of the former case.
	In detail, we set only one function whose wave packet covers non-oscillation subregion and continue to increase the resolution of the symbolic subregion by setting $M-1$ rectangles. Based on the above, the preprocessing process is complete.
	We thereby take KA method based on new generalized rectangle functions and correspondingly obtain the VLEZ as shown in Fig.\ref{fig:henon1_whole}(b).

	We find the coarse symbolic boundary subregion in the light of the oscillation and zero-valued subregions.
	Like one-dimensional map, the zero-valued regions can be assigned the same symbol as the adjacent oscillation subregion. Then, we find the symbolic region can be divided into three subregions based on continuity as the subplot shown in the Fig.\ref{fig:henon1_whole}(b). We refer to these subregions Part1, Part2 and Part3 from the left to right. We should take GKA method on three subregions separately.
	
	Firstly, we concentrate on the region Part1. We find that this region is similar to a curve and refer to this type of region as the curved-type region. 
	We find that the one-step evolved region of Part1, referred to as Part1S1, consisting of one incomplete folding and quite large non-folding subregion. Then we can remove the non-folding subregion of Part1S1 and the Part1 can also are shrunk as shown in Fig.\ref{fig:symbol henonA appd}(a). By shrinking, we can find that the Part1 is still curved-type region. Then, we continue to evolve Part1S1 and obtain Part1S2 and Part1S3. We find these regions are both curved-type regions. Thus, we can fit one curve of Part1 and other curve of Part1S2 or Part1S3 as the baselines for GKA method.

	Here, we find that coefficient of determination $R^{2}$ of Part1 is 0.9997 and $R^{2}$ of Part1S3  Part1S2 are 0.9999 and 0.9858 separately.
	However, we choose the complete folding region Part1S3 as the evolved region for KA method.
	Then we choose the two fitting curves as the baselines during the generalized affine transformation. Like one-dimensional case, we use the GKA method on this subregion. The steps of GKA methods on multivariate localized subregion are stated in detailed in the Appendix B.  
	We take GKA method with the parameters $M=20$, $n=381$ and $r_{w}=0.0377$ which is based on uniformly dividing the fitting curve of Part1S2 whose arc length $l=0.7154$. 
	and obtain the VLEZ. Obtaining the precise symbolic boundary of Part1, we take the GKA method for Part2 and similarly obtained the VLEZ as expected.
	
	However, after shrinking Part3 via removing the non-folding subregion of Part3S1, we find shrunk Part3 is not curved-type region($R^{2}<0.9$) but can be further separated into two curved-type subregions as the subplot shown in Fig.\ref{fig:symbol henonA appd}(b). We refer to the cyan region as Part3a and magenta region as Part3b. Then, we take GKA method on them separately. As presented in Fig.\ref{fig:symbol henonA appd}(b), the VLEZ of Part3a is obtained via taking Part3aS2 as evolved region and setting the parameters $M=20$, $n=381$ and $r_{w}=0.0596$.
	Similarly, we handle the Part3b and obtain the VLEZ.
	Finally, we refine whole symbolic boundary after obtaining the VLEZ from four symbolic subregions as is represented in Table \ref{tab:henon_boundary} and Fig.\ref{fig:henonB_whole}(d). 
	We take the scope of the zero-crossing points as symbolic boundary region shown in Table \ref{tab:henon_boundary}.

	When we obtain the relatively precise symbolic partition, it
	is necessary to compare this result with the effective partition conjectured by Grassberger and Kantz \cite{1985Generating}, given in Table \ref{tab:henon_boundary}.
	The symbolic boundary is nearly identical which justify the proposed method is applicative for some types of two-dimensional chaotic series.
	As a natural consequence, fractal structures are smeared out by our coarse graining,
	all branches of the attractor become "fat". However, we should cope with the noisy time series in that practical measured data contain noise.
	In the simulation of noisy time series, we add Gaussian white noise $\zeta_{n}$ and $\eta_{n}$ where the noisy strength $\sigma_{1}=\sigma_{2}=0.005$ into Eq.\eqref{eq:henon1_eq} and construct Eq.\eqref{eq:henon1_eq_sto} to generate a time series. 
	
	\begin{equation}
		\begin{split}
			x_{n+1} &=-\alpha x_{n}^{2}+y_{n}+1+\sigma_{1}\zeta_{n}, 	\\
			y_{n+1} &=\beta x_{n}+\sigma_{2}\eta_{n}
		\end{split}
		\label{eq:henon1_eq_sto}
	\end{equation}
	
	The symbolic partition of the noisy chaotic data can be obtained based on above algorithm. Fortunately, we find that the symbolic boundary is analogous to that without noise according to the presentation of the Table \ref{tab:henon_boundary}. 
	As is presented in the Table, we find that the symbolic boundary of the noisy chaotic data is precise via comparing with the that with $\sigma=0$. However, we cannot further refine in that noisy interference generates the inseparable small regions and disturb the accuracy of further refinement. 
	
	\begin{table*}%
		\caption{Comparison of obtained boundary and the theoretical boundary (given in \cite{1985Generating,2021Symbolic,1997Structure}) at different noisy level of the types of H\'{e}non map, where $\sigma$ = 0.005(H\'{e}non I) and 0.002(H\'{e}non II) for above two types of noisy series data respectively\label{tab:henon_boundary}}.
		\begin{tabular}{cclll}
			\cline{1-5}
			\multirow{2}{*}{Type}&\multirow{2}{*}{Local Subregion}& \multicolumn{2}{c}{practical boundary} &\multirow{2}{*}{theoretical boundary}\\
			\cline{3-4}
			&&  $\zeta_{n}=\eta_{n}=0$ & $\zeta_{n}=\eta_{n}=\sigma$\\
			\hline
			\multirow{4}{*}{h\'{e}non I}
			& Part1 &	$(0.704\pm 0.003, -0.002\pm 0.002)$  & 
			$(0.700\pm 0.020, -0.002\pm 0.002)$ & 
			( 0.703,-0.002) \\
			& Part2	&  	$(0.800\pm 0.000, 0.006\pm 0.001)$ & 
			$(0.800\pm 0.020, 0.006\pm 0.001)$ & 
			( 0.800, 0.006) \\
			& Part3a	&  $(1.231\pm 0.001, -0.025\pm 0.000)$  &  
			$(1.230\pm 0.010, -0.023\pm 0.002)$  & 
			
			( 1.231,-0.025)\\
			& Part3b	&  $(1.271\pm 0.001, -0.021\pm 0.000)$  & 
			$(1.270\pm 0.015, -0.019\pm 0.001)$ & 
			( 1.272,-0.021)\\
			\hline
			\multirow{6}{*}{h\'{e}non II}
			&Part1a &	$(-0.286\pm 0.003, 0.479\pm 0.000 )$ & 
			$(-0.290\pm 0.005, 0.473\pm 0.001 )$ &

			(-0.294, 0.478) \\
			& Part1b	&  	  $(-0.290\pm 0.003, 0.488\pm 0.001 )$ & 
			$(-0.290\pm 0.005, 0.492\pm 0.001 )$ & 
			(-0.293, 0.491) \\
			& Part2a	& $( 0.610\pm 0.000,-0.007\pm 0.003)$ & 
			$( 0.608\pm 0.000, -0.010\pm 0.005)$ & 
			( 0.609,-0.007)\\
			& Part2b	&   $( 0.648\pm 0.000, 0.008\pm 0.001)$ &  
			$( 0.645\pm 0.000, 0.008\pm 0.002)$ & 
			( 0.649, 0.009)\\
			& Part3a	& $( 1.584\pm 0.001, -0.100\pm 0.001)$ &
			$( 1.590\pm 0.010, -0.100\pm 0.002)$ & 
			( 1.581,-0.104)\\
			
			& Part3b	&  $( 1.610\pm 0.002, -0.094\pm 0.005)$ & 
			$( 1.620\pm 0.020, -0.081\pm 0.005)$ &
			( 1.622,-0.098)\\
			\hline
			
		\end{tabular}
	\end{table*}

	\subsubsection{series from H\'{e}non map II ($\alpha$=1.0, $\beta$=0.54)}
	\label{henonB}
	
	Compared to the previous case, here we deal with a situation
	in which the primary folding is extremely incomplete with the parameter value $\alpha=1.0$ and $\beta=5.4$ of Eq.\eqref{eq:henon1_eq}.
	In this condition, there are several symbolic partition results that yield equivalent symbolic dynamics descriptions but the boundaries of the symbolic partition are different\cite{1985Generating,hensethesis}.
	
	\begin{figure*}[!htbp] 
		
		\begin{minipage}[t]{0.48\linewidth}
			\centering
			\begin{subfigure}[t]{\textwidth}
				\centering
				\includegraphics[width=8.5cm]{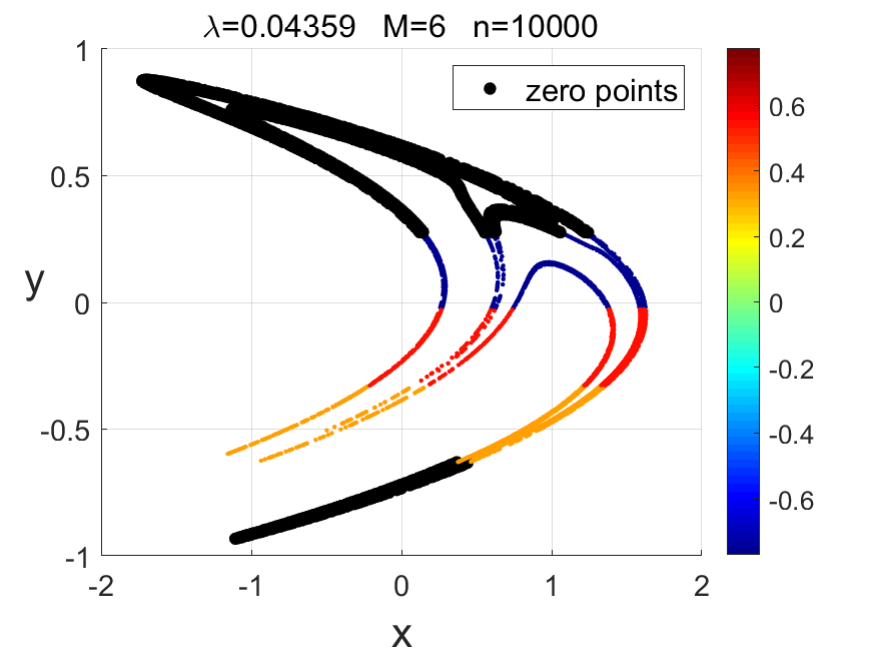}
				\caption{}
			\end{subfigure}
		\end{minipage}
		\hfill
		\begin{minipage}[t]{0.48\linewidth}
			\centering
			\begin{subfigure}[t]{\textwidth}
				\centering
				\includegraphics[width=8.5cm]{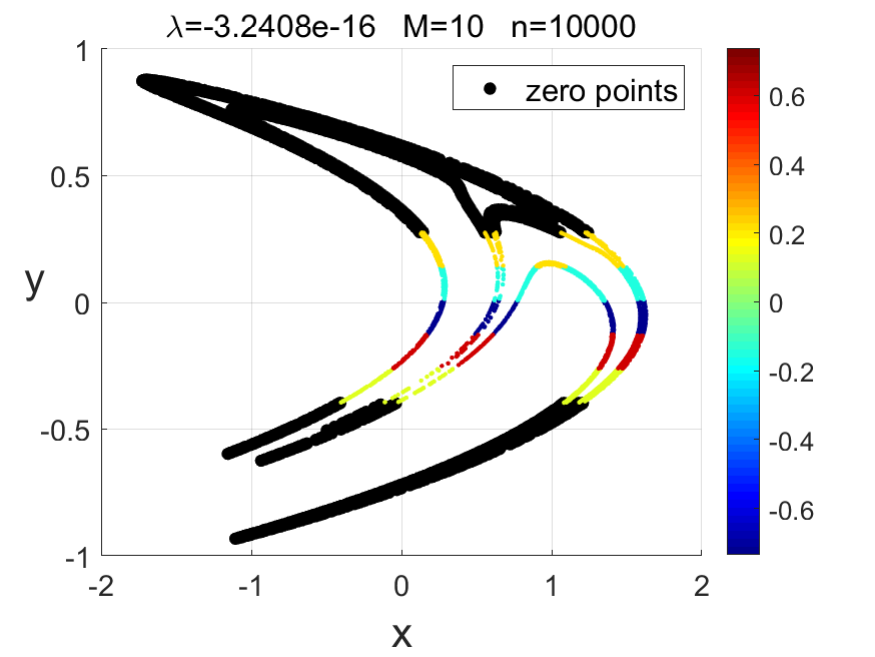}
				\caption{}
			\end{subfigure}
		\end{minipage}
		\hfill
		\begin{minipage}[t]{0.48\linewidth}
			\centering
			\begin{subfigure}[t]{\textwidth}
				\centering
				\includegraphics[width=8.5cm]{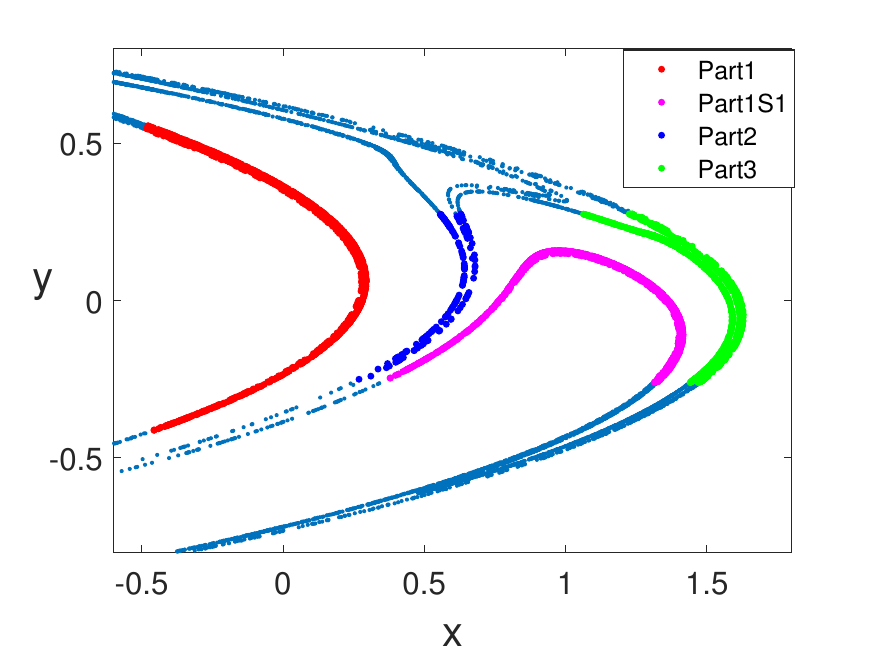}
				\caption{}
			\end{subfigure}
		\end{minipage}
		\hfill
		\begin{minipage}[t]{0.48\linewidth}
			\centering
			\begin{subfigure}[t]{\textwidth}
				\centering
				\includegraphics[width=8.5cm]{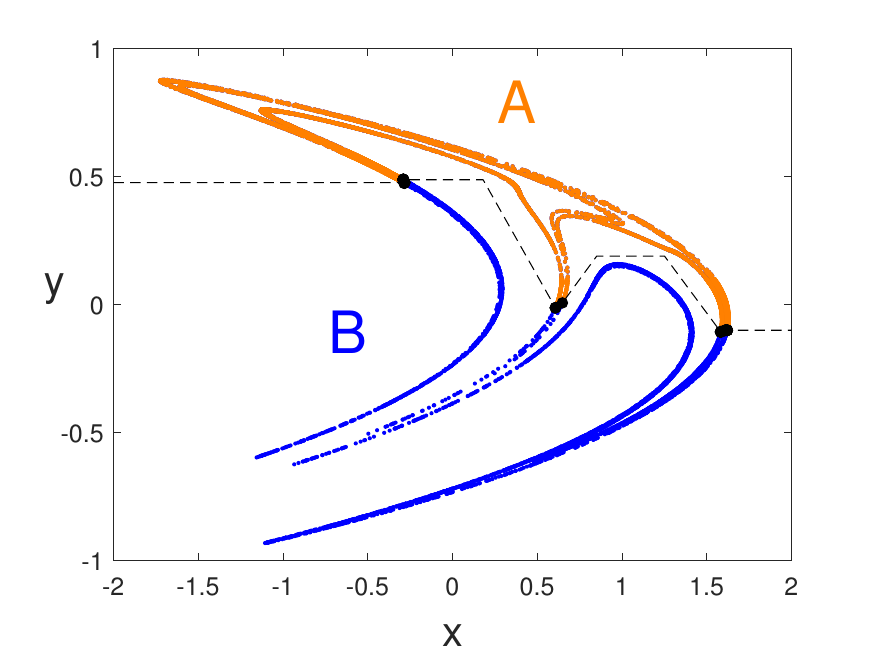}
				\caption{}
			\end{subfigure}
		\end{minipage}

		\caption{The KA method for the series governed by the H\'{e}non II map
			(a) The CLE via KA method for primary symbolic boundary.
			(b) The VLEZ via KA method for primary symbolic boundary.
			(c) The three modified subregions based on the coarse boundary.  
			(d)The eventually refined symbolic boundary
		}
		\label{fig:henonB_whole}
	\end{figure*}

	Like the previous example, we should construct the few rectangle basis functions whose wave packets cover the whole region. 
	Then, we chose two numbers of the segments of coordinate directions $x$ and $y$ are $Mx$ and $My$.
	Initially, we should ensure $Mx=My$. Then, we set the parameters are from $Mx=My=2$ gradually increasing $Mx=My$. Then, when $Mx=My\geq2$, we can find that the continuity of the fractal region cannot be broken when $Mx$ are reset to 1 considering the regularity of every subregion. Therefore, we chose $M=My$ and gradually increase $M$ from 2.
	We obtain the CLE when $M=6$ as shown in Fig.\ref{fig:henonB_whole}(a).
	We find that two zero-valued subregions. The below one is smaller than the above one and the above one is coarse folding region when we evolve the symbolic subregion.
	similarly, we redistribute the wave packets.
	Specifically, we assign one generalized rectangle wave packet into the above one.
	In addition, we assign $M-1$ wave packets into the rest region. When setting $M=10$, we can obtain MVLEZ as shown in Fig.\ref{fig:henonB_whole}(b).
	We neglect light green part of the symbolic region in that they are similar to zero-valued region and obtain four continuous subregions. We find that the leftmost region and the second region on the right are the relationship of forward and backward region. Based on incomplete folding, we should choose one subregion as the symbolic region and the other as the zero-valued region. Based on the continuity of whole region, we retain the leftmost region and removing the other.
	As is shown in Fig.\ref{fig:henonB_whole}(c), we retain the leftmost region and enlarge it based on the backward region of the second region on the right and obtain the Part1. The other region is naturally referred to as Part1S1. The rest two subregions are Part2 and Part3.
	We take GKA method on Part1 and obtain VLEZ. Then we further refine symbolic boundary via separating the localized Part1 into two curved-type subregions Part1a and Part1b. Then we obtain the VLEZ via GKA method and accordingly obtain precise symbolic boundaries as presented in Table \ref{tab:henon_boundary}.

	Similarly, we shrink Part2 via removing redundant part of Part2S2. We obtain two curved-type subregions Part2a and Part2b as shown in Fig.\ref{fig:symbol henonB appd}(a).
	Here, we start to handle Part2a.
	However, we find Part2 do not have enough state points, with only 27 points instead of the required 100.
	In this condition, we should take interpolation increases the state points. specifically, we
	fit curve of it as the red curve where $R^{2}=0.9997$ and set $n=100$ points with equal arc length on the curve as interpolation points. Due the $R^{2}>0.99$ , we can just take interpolation points as state points instead of adding to state points. Then, we find
	multiple-step evolved region Part2aS4 and fit straight line with $R^{2}=0.9910$ as the below subplot in Fig.\ref{fig:symbol henonB appd}(a). It is worth noting that we fit line rather than curve in that we need to construct the one-dimensional localized evolution map by fitting two-dimensional curve regarding the original points after generalized affine transformation and evovled original points as the above subplot in Fig.\ref{fig:symbol henonB appd}(a).
	Then, we take the 100 interpolation points as the independent variable and obtained 100 evolved state points based on the evolution map.
	We fitting the map based on only one coordinate direction. Therefore, we have to ensure that this evolution map is suitable for all coordinate directions and thus generalize whole two-dimensional region. Base on this, we should ensure that the evolved region need to fitting straight line. Only in this way, evolved state points based on the fitting evolution map are appropriate and we can take GKA method for interpolation points.
	Then, we take GKA method with parameters $M=20$, $n=100$ and $r_{w}=0.0016$ and obtain the VLEZ as the subplot shown in Fig.\ref{fig:symbol henonB appd}(b). However, the boundary position is not highly correct because the $R^{2}$ of fitting straight line of Part2aS4 is 0.9910 and thus is not exactly similar to two-dimensional curve. Therefore, we continue to localize the Part2a based on the oscillation of the original VLEZ. Thus, we localized the original points of Part2a with the scope $[-0.015,0.03]$ and fitting straight line of Part2aS4 where $R^{2}=1.0000$ and then we obtain the VLEZ in Fig.\ref{fig:symbol henonB appd}(b) and thus the precise symbolic boundary.
	Then, we take GKA method on Part3 and obtain the narrow symbolic region. Similarly, the shrunk Part3 can be separated into two curved-typed subregions, namely Part3a and Part3b.

	After refining the primary coarse boundary via separating it into six subregions, namely Part1a, Part1b, Part2a, Part2b, Part3a and Part3b. We take GKA method on them respectively and obtain the symbolic boundaries as represented in Table \ref{tab:henon_boundary} and Fig.\ref{fig:henonB_whole}(d).

	We successfully apply our approach to the chaotic series based on the chaotic system with strongly incomplete folding after comparing the result with one type of accurate symbolic boundary position obtained by Hansen \cite{hensethesis}.
	We add small Gaussian white noise with the noisy strength $\sigma=0.002$ to the chaotic data and obtain the similar partition result by the above algorithm as presented in Table\ref{tab:henon_boundary}.

	\subsubsection{series from Duffing oscillator }
	
	In Section\ref{henonA} and Section\ref{henonB}, we deal with classical discrete chaotic systems, indicating that our approach is suitable for discrete chaotic systems. However, real signal may be continuous, from autonomous and non-autonomous systems. We can discretize them and apply Poincaré sections to our systems, achieving both discretization and reduction of the time dimensionality, thereby completing data preprocessing. 
	
	\begin{figure*}[!htbp] 
		\begin{minipage}[t]{0.48\linewidth}
			\centering
			\begin{subfigure}[t]{\textwidth}
				\centering
				\includegraphics[width=8.5cm]{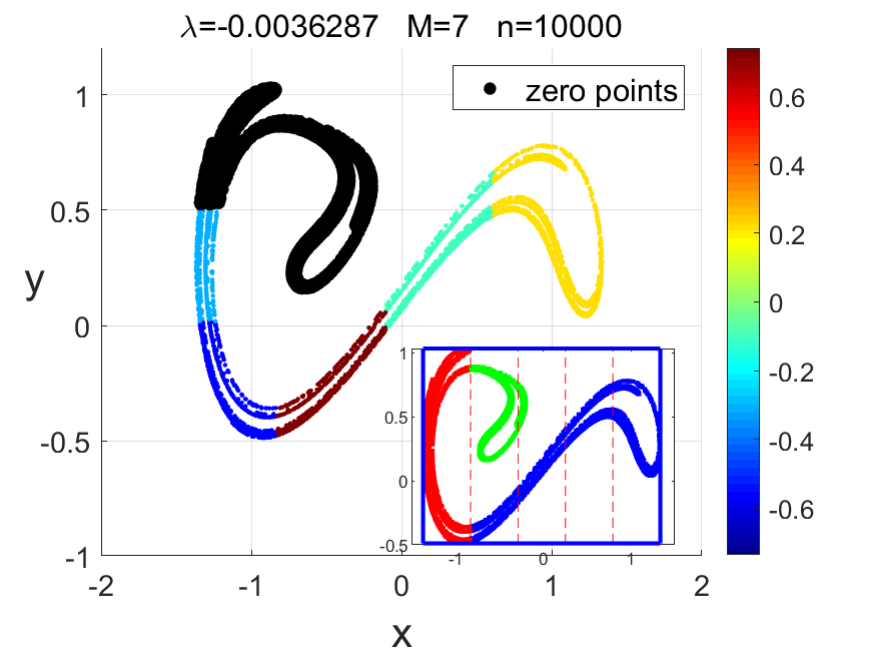}
				\caption{}
			\end{subfigure}
		\end{minipage}
		\hfill
		\begin{minipage}[t]{0.48\linewidth}
			\centering
			\begin{subfigure}[t]{\textwidth}
				\centering
				\includegraphics[width=8.5cm]{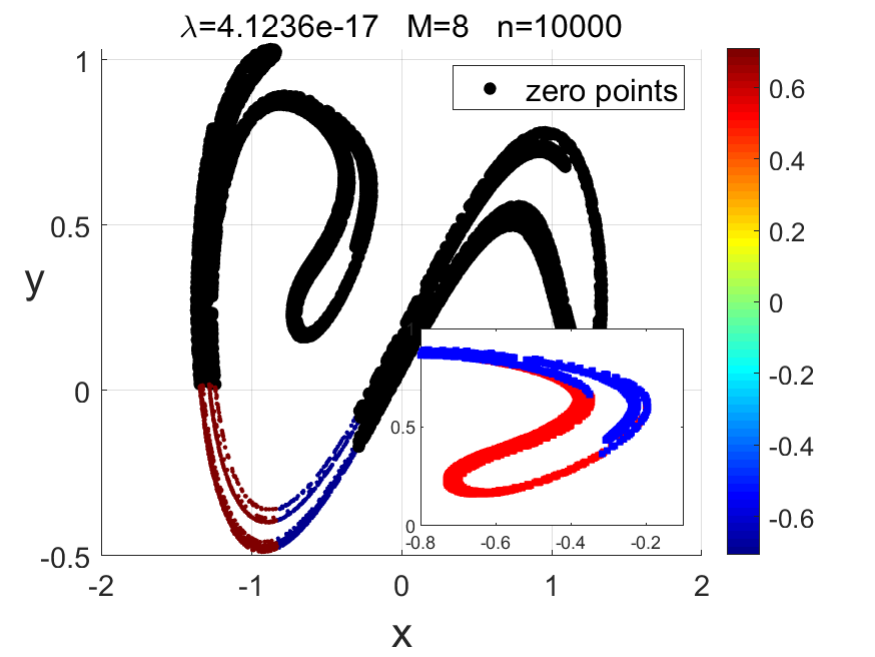}
				\caption{}
			\end{subfigure}
		\end{minipage}
		\hfill
		\begin{minipage}[t]{0.48\linewidth}
			\centering
			\begin{subfigure}[t]{\textwidth}
				\centering
				\includegraphics[width=8.5cm]{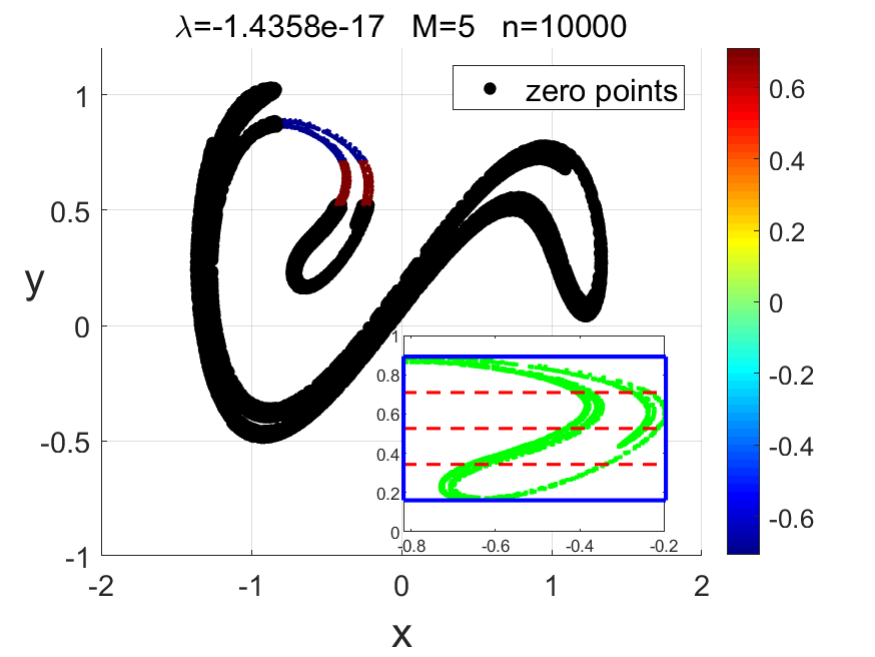}
				\caption{}
			\end{subfigure}
		\end{minipage}
		\hfill
		\begin{minipage}[t]{0.48\linewidth}
			\centering
			\begin{subfigure}[t]{\textwidth}
				\centering
				\includegraphics[width=8.5cm]{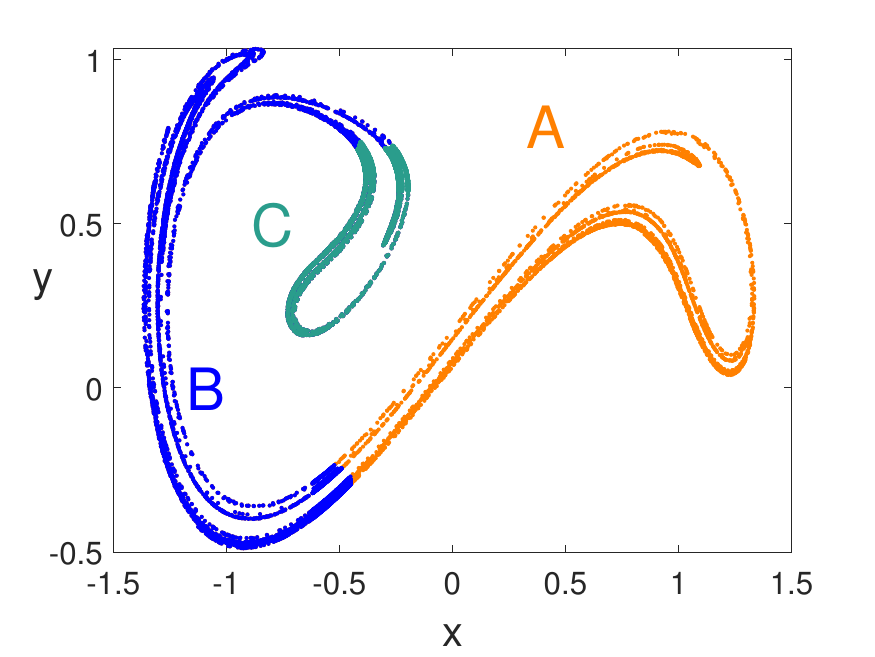}
				\caption{}
			\end{subfigure}
		\end{minipage}

		\caption{The symbolic partition of the series from the Duffing oscillator return map Eq.\eqref{eq:duffing} via KA method.
			(a) The CLE via KA method for primary symbolic boundary. 
			(b)-(c) The VLEZ via KA method for different two primary symbolic boundaries.
			(d) Eventual refined symbolic boundaries.
		}
		\label{fig:koopman_part_duffing}
	\end{figure*}
	
	We partition the time series from the Duffing oscillator as a challenging example of the non-autonomous system. After the data preprocessing by periodic sampling, the return map of the Duffing oscillator governed by 
	
	\begin{equation}
		\begin{split}
			\ddot{x}+\alpha\dot{x}-x+x^{3} = \beta cos t
		\end{split}
		\label{eq:duffing}
	\end{equation}
	where the parameters are $\alpha=0.25$ and $\beta=0.4$. This chaotic series under the above parameters was firstly partitioned into different symbolic regions by F. Giovannini and A. Politi\cite{1991Homoclinic}.
	Obtaining the series from the return map of Eq.\eqref{eq:duffing}, we can find the region in two-dimensional state space. So we can use above analogous pretreatment steps and proposed methods to handle the symbolic partition. 
	We take the KA method to cope with this chaotic time series similar as those of Section \ref{henonA} and Section \ref{henonB} to obtain a primary boundary.

	Based on the continuity, we initially set the parameters are from $Mx=My=2$ gradually increasing $Mx=My$. When $Mx=My=2$, we can find that the continuity of the fractal region cannot be broken when $My$ are reset to 1 considering the regularity of every subregion.
	
	Different with the above two maps, when $Mx=My\geq3$, we can find that the continuity of the fractal region and the regularity of every subregion are both broken when $My$ to 1. 
	Therefore, we cannot divide it into continuous subregions based on constructing many rectangles from only one coordinate direction, which is an obstacle to pretreatment process.
	To tackle this issue effectively, it’s essential to divide original region into multiple subregions where every region can further be divided into continuous subregions based on only one coordinate direction separately. Specifically, we increase $Mx$ to 5 and redistribute these subregions into new three subregions are shown in the subplot in Fig.\ref{fig:koopman_part_duffing}(a).
	We refer to the blue, red and green subregions as Part$\alpha$, Part$\beta$ and Part$\gamma$. we construct equal rectangles respectively for three subregions based on their distinctive coordinate direction. However, we should ensure the rectangle packets from three subregions are approximately same. Here, we ensure that the ratios between pairs of the three kinds of packet primary edge lengths are all within $[0.8,1]$. Then we construct the generalized rectangle basis functions when set $M1=M1x=3$ packets for Part$\alpha$, $M2=M2y=3$ packets for Part$\beta$ and $M3=M3y=1$ packets for Part$\gamma$. Here, we construct $M=M1+M2+M3=7$ basis functions and accordingly obtain the CLE by KA method as shown in Fig.\ref{fig:koopman_part_duffing}(a). We find the two connected zero-valued subregions and retail marginal one, Part$\gamma$, as the non-symbolic region. Like above cases, we should perform the process of redistribution. Specifically, we assign one wave packet on this region and the rest regions Part$\alpha$, Part$\beta$ should be assigned $M-1$ packets.
	Here, we set $M1=4$ packets for Part$\alpha$ and $M2=3$ packets for Part$\beta$ and obtain UVLEZ and primary symbolic subregion Part1 by KA method in Fig.\ref{fig:koopman_part_duffing}(b). 
	However, we find that the one-step evolved region Part1S1, as the red region shown in the subplot of Fig.\ref{fig:koopman_part_duffing}(b), is a special folding where the complete overlap position is not in the top of region. Then, we should continue to evolve Part1S1 and obtain Part1S2. Then, we shrink the Part1S2 and thus also shrink Part1 and obtain two curved-type region Part1a and Part1b. Next, we take GKA method and interpolation on two surbregions respectivley and accordingly obtain the refine the symbolic boundary. Similarly, we further refine them via set Part1aS3 and Part1bS3 in place of Part1aS2 and Part1bS2 respectivley.

	After obtaining the first boundary, we find that it is possible to have other boundaries for this special folding. We consider Part$\gamma$ as the first folding region via above analysis. However, the pre-image of Part$\gamma$ is not only the two surrounding subregions of the first symbolic boundary but also the partial subregion of Part$\gamma$, which indicates that the first folding region Part$\gamma$ consist three subregions and consequently we should find the third symbol subregion and second boundary.
	The coarse second boundary is Part$\gamma$ based on above analysis. Then, we should refine the coarse second boundary. Likewise, we go through the process of redistribution again. As the subplot in Fig.\ref{fig:koopman_part_duffing}(c), we divide Part$\gamma$ into $M-1$ rectangles based on $y$ coordinate. The rest part covers only one packet. Then, we take KA method and obtain VLEZ just when $M=4$. Then, we increase $M$ to 5 and obtain the VLEZ where the oscillation is narrow than the former VLEZ. We can take GKA method to refine the second boundary when we further shrink the oscillation subregion in Fig.\ref{fig:koopman_part_duffing}(c) and separate it into two subregions Part2a and Part2b.

	After refinement of two symbolic subregions, we finish the eventually symbolic boundaries as shown in Fig.\ref{fig:koopman_part_duffing} (d).  
	The eventually symbolic partition we obtained is similar to the former results\cite{2018Empirical,1991Homoclinic}.
	Successful symbolic partition for this time series indicates that our method can successfully divide non-autonomous chaotic systems into different symbols and further demonstrates the strong universality of this method. Certainly, the noise test has also been successful in such chaotic series.

	\subsubsection{series from three-dimensional hyperchaotic chaotic map}
	\label{threedim}
	
	In this section, we will extend its application to a three-dimensional hyperchaotic map, which is proposed by Baier and Klein \cite{generalized}, described as
	
	\begin{equation}
		\begin{aligned}
			x_{n+1} &= \alpha-y_{n}^{2}+\beta z_{n},\\
			y_{n+1} &= x_{n},\\
			z_{n+1} &= y_{n},\\
		\end{aligned}
		\label{eq:three_dim}
	\end{equation}
	where $\alpha = 1.76$ and $\beta = 0.1$.
	
	\begin{figure*}[!htbp] 
		\begin{minipage}[t]{0.48\linewidth}
			\centering
			\begin{subfigure}[t]{\textwidth}
				\centering
				\includegraphics[width=8.5cm]{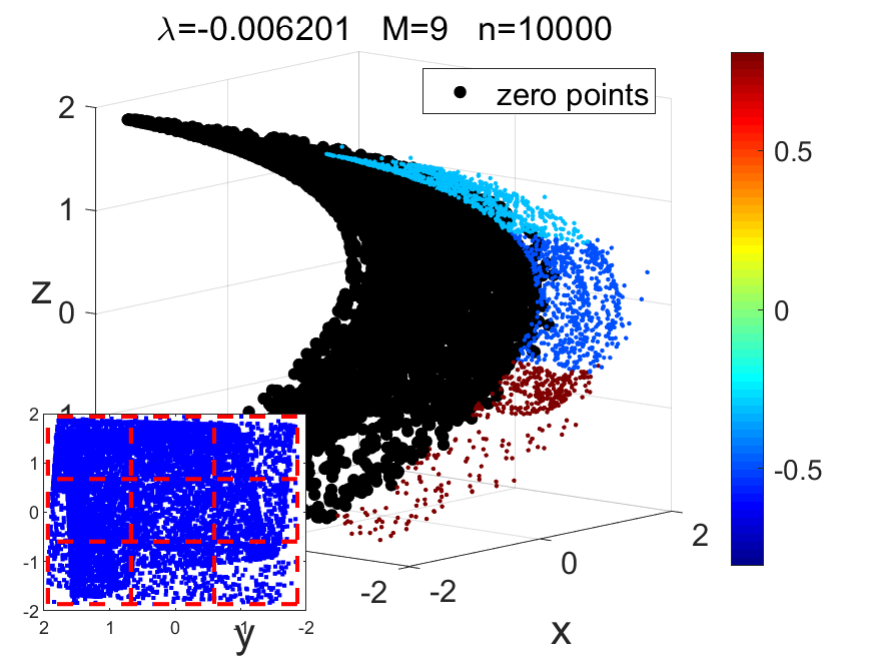}
				\caption{}
			\end{subfigure}
		\end{minipage}
		\hfill
		\begin{minipage}[t]{0.48\linewidth}
			\centering
			\begin{subfigure}[t]{\textwidth}
				\centering
				\includegraphics[width=8.5cm]{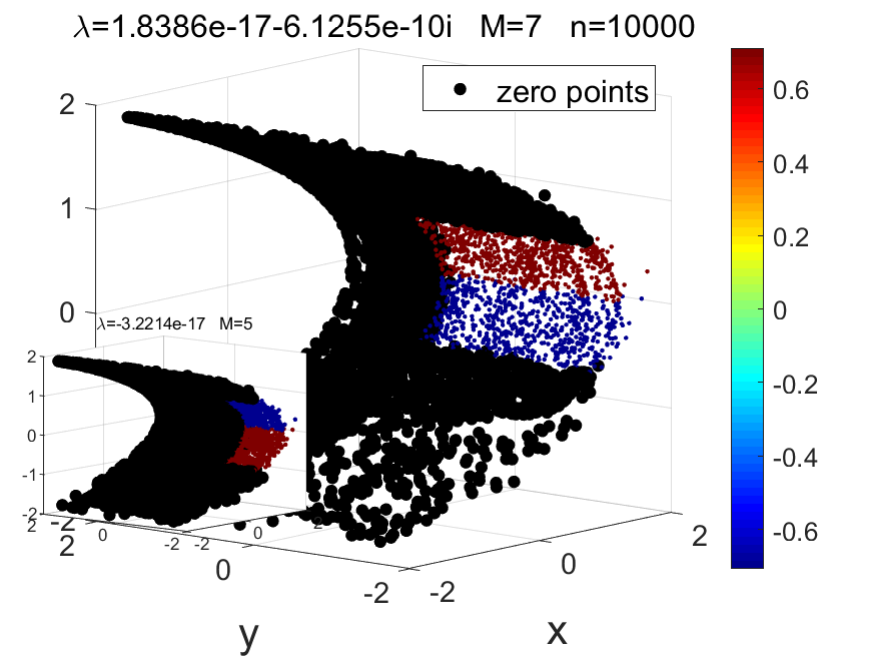}
				\caption{}
			\end{subfigure}
		\end{minipage}
		\hfill
		\begin{minipage}[t]{0.48\linewidth}
			\centering
			\begin{subfigure}[t]{\textwidth}
				\centering
				\includegraphics[width=8.5cm]{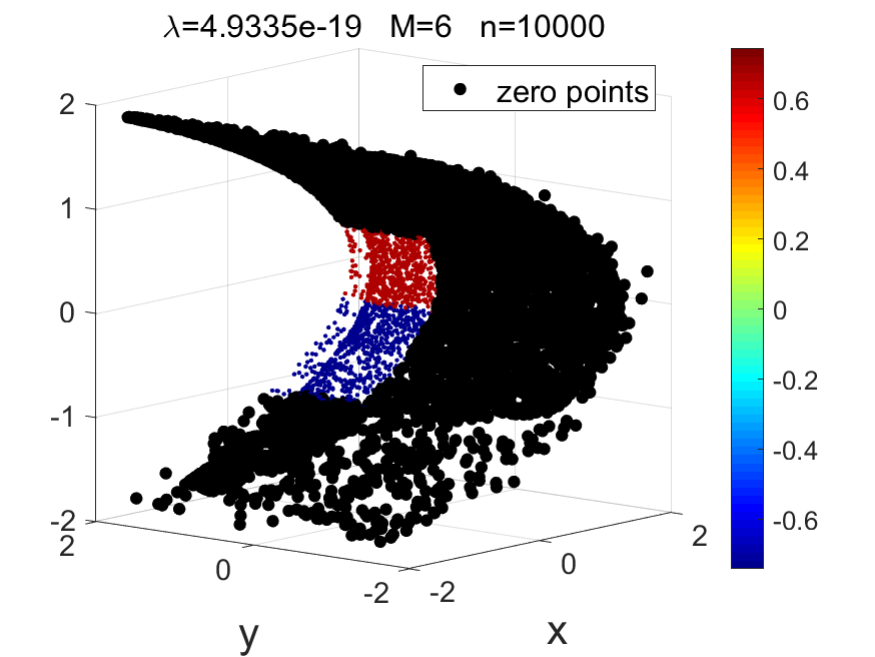}
				\caption{}
			\end{subfigure}
		\end{minipage}
		\hfill
		\begin{minipage}[t]{0.48\linewidth}
			\centering
			\begin{subfigure}[t]{\textwidth}
				\centering
				\includegraphics[width=8.5cm]{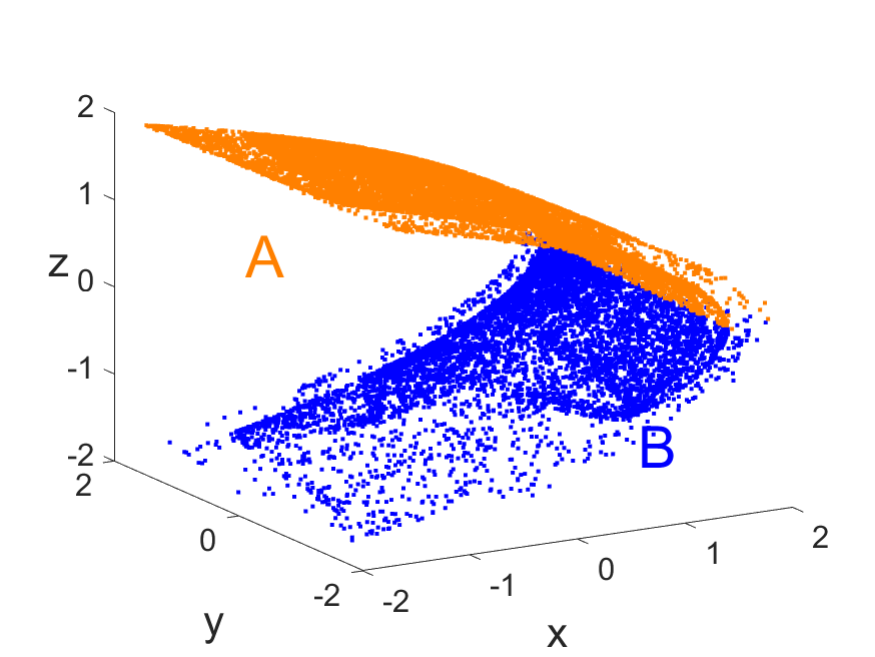}
				\caption{}
			\end{subfigure}
		\end{minipage}
		
		\caption{The symbolic partition of the 3-dimensional series from the map Eq.\eqref{eq:three_dim} 
			(a) The CLE via KA method for primary symbolic boundary. 
			(b)-(c) The VLEZ to obtain different two primary symbolic boundaries
			(d) Eventual refined symbolic boundary
		}
		\label{fig:three_dim}
	\end{figure*} 
	At the current parameter values, the system is chaotic and the attractor strcture is much more complicated than that of bi-variable chaotic series.
	Similarly to above examples, we need to obtain the primary boundary of the whole region which consists of $n=10000$ chaotic state points. We still use the KA method and GKA method for this case.
	Firstly, we construct a three-dimensional rectangular cuboid region whose boundaries are all also tangent with the attractor just like the steps for the above two-dimensional maps. 
	Then we uniformly separate the whole rectangular cuboids in each coordinate direction into many uniform small rectangular cuboids as initial wave packet regions. 
	The three numbers of the segments of coordinate directions $x$, $y$ and $z$ are $Mx$, $My$ and $Mz$.
	Initially, we should ensure $Mx=My=Mz$. Then, we set the parameters are $Mx=My=2$ and obtain the $2\times2\times2$ three-variate generalized rectangular cuboids. At that parameter, we can find that the continuity of the fractal region cannot be broken when $Mx$ are reset to 1 where we also ensure the regularity of every subregion. When $Mx=My=Mz\geq2$, $Mx$ should be reset to 1 based on the same cause.
	Therefore, we chose $M= My\times Mz$ and gradually increase $My/Mz$ from 2.
	we obtain the $3\times3$ three-variate generalized rectangle functions whose wave packets are all continuous and the projection onto the y and z axes as the subplot in Fig.\ref{fig:three_dim}(a).
	Thus we take KA method with the parameters set as $M=9$ and obtain the CLE as is shown in Fig.\ref{fig:three_dim}(a).
	We find that oscillation appears in right three subregions among all nine subregions from the view of y coordinate. Then, the CLE demonstrates that the boundary line almost primarily in parallel with coordinates $y$ in this case. 
	As a result, we should divide the region into the segments with only $z$ coordinate direction. Therefore, we can further set $M=Mz$ in the following text.
	
	Then, we improve the resolution of the oscillation region with $M-1=4$ wave packets.
	The zero-valued region consisting of other six subregions are covered with only one wave packet. After conducting above pretreatment process, we subsequently take KA method based on the updated $M=5$ subregions and obtain the VLEZ as the subplot in Fig.\ref{fig:three_dim}(b). However, we find that the zero-valued region is so large that we should shrink it. Then, we return to the $3\times3$ three-variate generalized rectangular cuboids and take right six subregions as oscillation region while left three subregions as nonoscillation region where only one packet covers it. We gradually improve the resolution of the oscillation region and obtain the VLEZ when $M-1=6$ as shown in Fig.\ref{fig:three_dim}(b). Here, we find that the eigenvalue contains an imaginary part, indicating that a significant rotation occurred during the folding process. Fortunately, 
	the module of the eigenvalue still minimal among $M$ eigenvalues and the error is within $10^{-5}$ of zero.
	
	Then, we merge rest regions into one symbolic region and thus divide it into $M-1$ cuboids. Then, we further take KA method and obtain VLEZ when $M=7$ shown in Fig.\ref{fig:three_dim}(b). Here, we find that the boundary still does not partition the left zero-valued region when considering connect between symbolic region and non-symbolic region. 
	Thus, we should take the left zero-valued region as the symbolic region and take $M-1=5$ rectangular cuboids as packets of basis function. Meanwhile, we cover just one packet on the rest region. When $M=6$, we obtain the VLEZ shown in Fig.\ref{fig:three_dim}(c).

	After obtaining the coarse boundary, then we refine the primary boundary via the separation of localized symbolic subregion and GKA method as described in Appendix B. The final precise symbolic partition of the series from the map Eq.\eqref{eq:three_dim} is presented in Fig.\ref{fig:three_dim}(d).
	Noisy series from Eq.\eqref{eq:three_dim} can also achieve almost similar partition positions, demonstrating that our approach can be extended to complex high dimensional hyperchaotic map.
	
	\section{Conclusion}
	\label{sec:conclusion}
	
	In conclusion, our methods integrate Koopman theory to handle symbolic partitioning of chaotic time series in that we discover the relationship between Koopman left eigenfunction and folding behavior of chaotic series point set evolution. To certificate the effectiveness of this method, we introduce one-dimensional unimodal map chaotic data as a case study to explain the underlying mechanism of our approach because the foldings of one-dimensional chaotic maps are complete, which is the precise location of the symbolic boundary. 
	In addition, when we simplify the computation of obtaining VLEZ by localizing the whole region, we improve the KA method into GKA method, which combines affine transformation with the KA method to construct an equivalent unimodal map. In addition, these two methods are also applied to series from multimodal map and even the multivariate chaotic map. Furthermore, after demonstrating the feasibility of our proposed method, we extend our method to complex applications. 
	
	We generalize methods to the multimodal univariate map and multivariate chaotic map by their features. We encounter difficulties in the incomplete primary folding and interference from other foldings. To address this, we take two steps for primary boundary. We initially obtain CLE and construct redistributed new basis functions via enhancing resolution of the estimated boundary and reducing the resolution of other subregions simultaneously based on the result of the CLE. Then, we further take KA method to obtain VLEZ.
	After obtaining primary boundary for multivariate chaotic map, we should divide the localized boundary region into multiple curved-type subregions and perform the GKA method on each of the subregions. 
	Specifically, we perform symbol partition on the series from two types of H\'{e}non maps and the return map of Duffing oscillator, as well as three-dimensional chaotic maps data even in the presence of weak observation noise. 
	The advantages of using this method are manifold: it is capable of handling complex series from high-dimensional and continuous systems, demonstrates a certain degree of robustness, and is meaningful for processing real-world data. This approach not only expands the methods available for symbolic partitioning but also extends the application scope of Koopman operator theory.
	A comparison with other methods to identify the symbolic boundary has been carried out showing some similarities and differences. Firstly, similar to the method based on unstable manifold \cite{2002chaos} in the first category, it applies the principle of stretching and folding. However, the distinction lies in its lack of requirement to consider the initial fixed point of evolution and its ability to accommodate a certain level of noise while those methods hardly handle this.
	Secondly, similar to most methods of the second category, all of them proceed in two steps: initially, roughly determining the positions of the symbolic boundary, followed by precise adjustments to their placement. However, the difference is that coarse primary boundary of ours is determined based on the principle of stretching and folding while the others are generally ambiguous. Finally, despite utilizing Koopman operator theory, yet, Cong Zhang's approach\cite{2022Phase} evidently lacks the advantages of the aforementioned methods from above two categories. It is applicable to  a limited set of simpler chaotic time series, whereas our method can handle the series from a diverse range of highly complex chaotic systems.
	However, our method is mainly applied to discrete chaotic series at present. In other words, many issues in ordinary continuous systems still retain when we apply our method to them. 
	Hence, it is essential to improve this method to more general chaotic systems. These are areas that require further in-depth research and exploration.
	
	In summary, we present a practical algorithm
	to estimate and refine symbolic boundary from time series
	from diverse chaotic systems. It addresses  problems  
	from operator-free methods and is feasible for realistic time series.

	\section*{Acknowledgements}
	This work was supported by the National Natural Science Foundation of
	China under Grants No.12375030.
	
	\appendix 

	\section*{Appendix A  Representation of different types of basis functions }
	
	\setcounter{equation}{0} 
	
	\renewcommand{\theequation}{A.\arabic{equation}}

	We take the following three types of functions as the basis functions in this article.
	The rectangle function is defined
	by
	\begin{equation}
		g_{Rj}(x)=
		\begin{cases}
			1,\ &(\dfrac{j-1}{M}l(\mathbb{R}^d)\leqslant x<\dfrac{j}{M}l(\mathbb{R}^d))\\
			0,\ &(\mbox{otherwise})
		\end{cases},\ j=1,2,\cdots,M\\
		\label{eq_rect}
	\end{equation}
	where  $l(\mathbb{R}^d)$ denotes the length of the one-dimensional region $\mathbb{R}^d$. The function Eq.\eqref{eq_rect}
	is used for one-dimensional chaotic series in the spectral decomposition of the Koopman operator in this paper.
	
	Considering the limitation of the function Eq.\eqref{eq_rect}, we should generalize the rectangle function for handle the multivariate chaotic series
	and the function is
	\begin{equation}
		g_{GRj}(\mathbf{x})=
		\begin{cases}
			1,\ &( \mathbf{x}\in S_{j})\\
			0,\ &(\mbox{otherwise})
		\end{cases},\ j=1,2,\cdots,M\\
		\label{eq_gener_rect}
	\end{equation}
	where $S_{j}$ is an ordinary localized region which might be structurally irregular subregion, instead of a uniform and regular subregion just as a line segment, rectangle or cuboid. Here, we refer to this function as generalized rectangle function in that the other parts are the same as a standard rectangle function.

	In addition to the above two functions, we can choose Gaussian function to handle the one-dimensional cases and multivariate curved-type region. The Gaussian function is defined

	\begin{equation}
		g_{GAi}(\mathbf{x}) = \mathbf{exp}(-\dfrac{| \mathbf{x}-\mathbf{x}^{*}_{j}|^2}{r_{w}^{2}}),j=1,2,...,M
		\label{gauss}
	\end{equation}
	where the $\mathbf{x^{*}}$ is the function grid point at the center of each wave packet and $r_{w}$ is the radius width of wave packet of the basis functions.
	We chose $\mathbf{x^{*}}$ based on the principle that grid points uniformly distributed as much as possible across the entire region. In addition, a wave packet should cover the region whose two borders are the two adjacent grid points from one grid point. Therefore, the radius of the wave packet $r_{w}$ is the distance between two grid points. In the case of the univariate chaotic regions, we assign $M$ uniform grid points as 
	
	\begin{equation}
		\mathbf{x^{*}_{\text{j}}}=\frac{l(\mathbb{R}^d)}{M}j+\min(\mathbf{x}), j=1,2,\ldots,M
		\label{eq:xj}
	\end{equation}
	where $l(\mathbb{R}^d)$ denotes length of the linear region or arc length of the curved-type region $\mathbb{R}^d$. 
	The other parameter $r_{w}$ is 
	\begin{equation}
		r_{w}=\frac{l(\mathbb{R}^d)}{M}
		\label{eq:xrw}
	\end{equation}
	where $r_{w}$ is the distance between two adjacent grid points and thus can be regarded as the radius of the Gaussian wave packet in this article. 
	
	Considering the efficient convergence of the result of the accurate symbolic boundary, we usually take this function Eq.\eqref{gauss} to handle the localized subregions for refining the boundary. Therefore, the basis function Eq.\eqref{gauss} is always appplied to GKA methods.
	
	\section*{Appendix B  Related Details Explanation for GKA method}
	
	\setcounter{equation}{0}
	
	\setcounter{figure}{0}
	
	\renewcommand{\theequation}{B.\arabic{equation}} 
	
	\renewcommand{\thefigure}{B.\arabic{figure}}  
	
\begin{figure*}[!htbp] 
		\begin{minipage}[t]{0.48\linewidth}
			\centering
			\begin{subfigure}[t]{\textwidth}
				\centering
				\includegraphics[width=8.5cm]{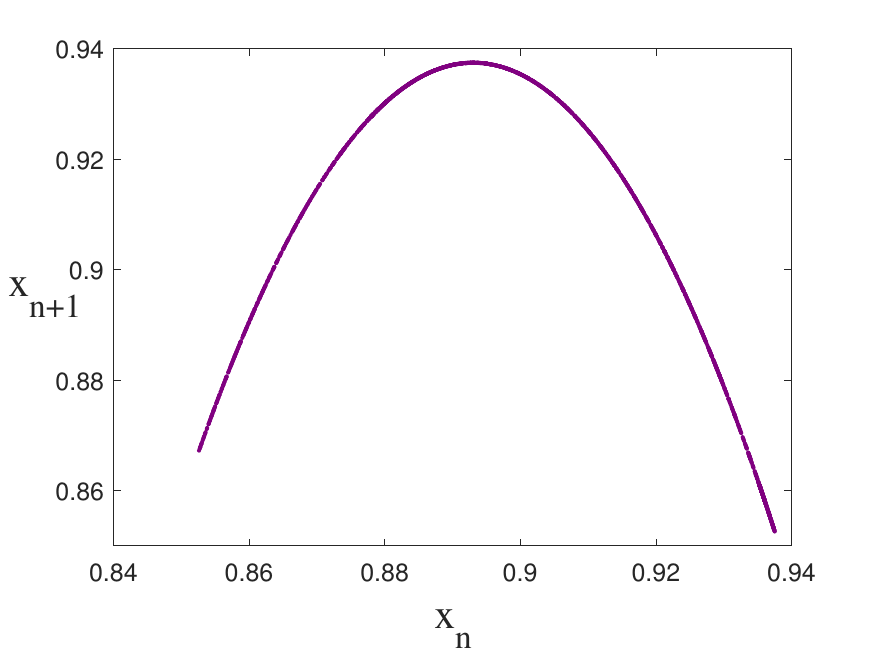}
				\caption{}
			\end{subfigure}
		\end{minipage}
		\hfill
		\begin{minipage}[t]{0.48\linewidth}
			\centering
			\begin{subfigure}[t]{\textwidth}
				\centering
				\includegraphics[width=8.5cm]{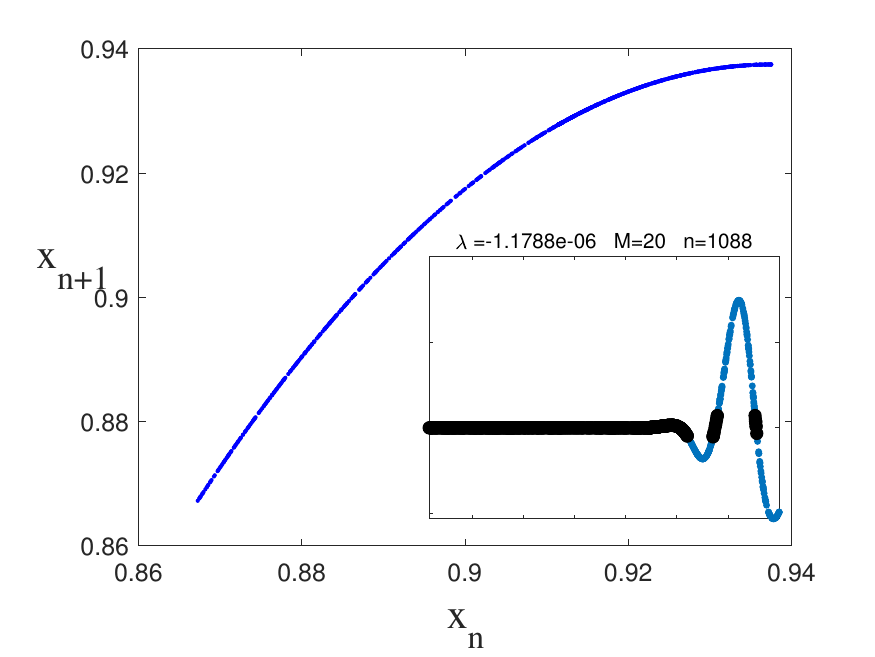}
				\caption{}
			\end{subfigure}
		\end{minipage}
		\caption{
			(a) The graph relating $\mathbf{x_{\text{aff}}}$ to $\mathbf{x_{\text{p}}}$ 		 
			(b) The localized non-chaotic map and one invalid LEZ via KA method
		}
		\label{fig:symbol one appd}
	\end{figure*}

	We partition the logistic map and obtain the VLEZ at low resolution for identifying the primary symbolic subregion. For refining primary symbolic subregion, we take GKA method. Here, the evolution relation between $\mathbf{x_{\text{aff}}}$ and $\mathbf{x_{\text{p}}}$ as a supplement for explanation for GKA method on one-dimensional chaotic map as presented in Fig.\ref{fig:symbol one appd}. 
	
	We always fit polynomial curve where the highest term is less than or equal to five times for curved-type regions which are always the localized subregions of the whole chaotic attractor. Because one curve can be transferred into another curve, so we take this step for two original regions.

	\begin{figure*}[!htbp] 
		\begin{minipage}[t]{0.48\linewidth}
			\centering
			\begin{subfigure}[t]{\textwidth}
				\centering
				\includegraphics[width=8.5cm]{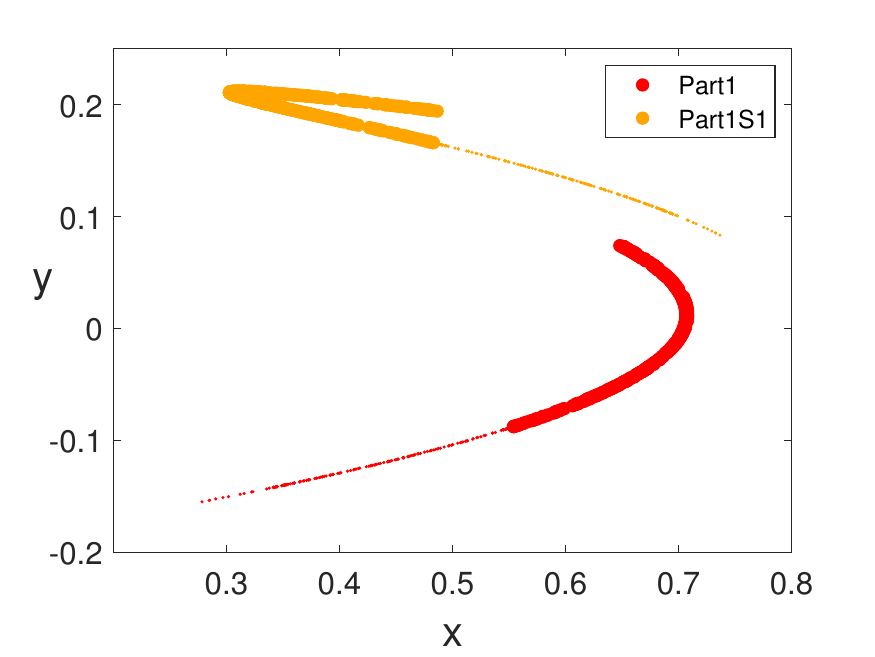}
				\caption{}
			\end{subfigure}
		\end{minipage}
		\hfill
		\begin{minipage}[t]{0.48\linewidth}
			\centering
			\begin{subfigure}[t]{\textwidth}
				\centering
				\includegraphics[width=8.5cm]{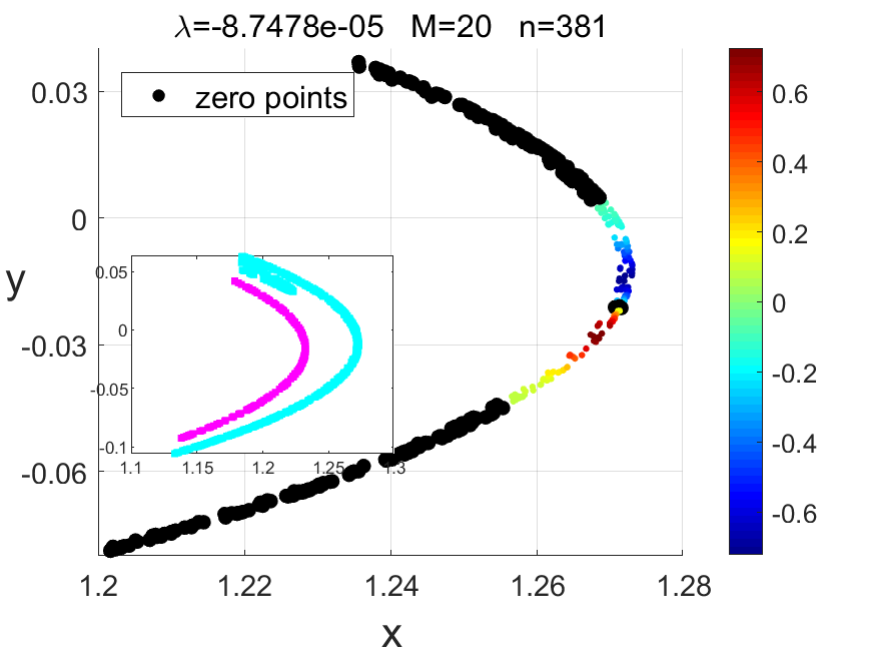}
				\caption{}
			\end{subfigure}
		\end{minipage}
		\caption{
			(a)  The shrunk Part1 and Part1S1
			(b) The VLEZ via GKA method for Part3a
		}
		\label{fig:symbol henonA appd}
	\end{figure*}
	%

	\begin{figure*}[!htbp] 
		\begin{minipage}[t]{0.48\linewidth}
			\centering
			\begin{subfigure}[t]{\textwidth}
				\centering
				\includegraphics[width=8.5cm]{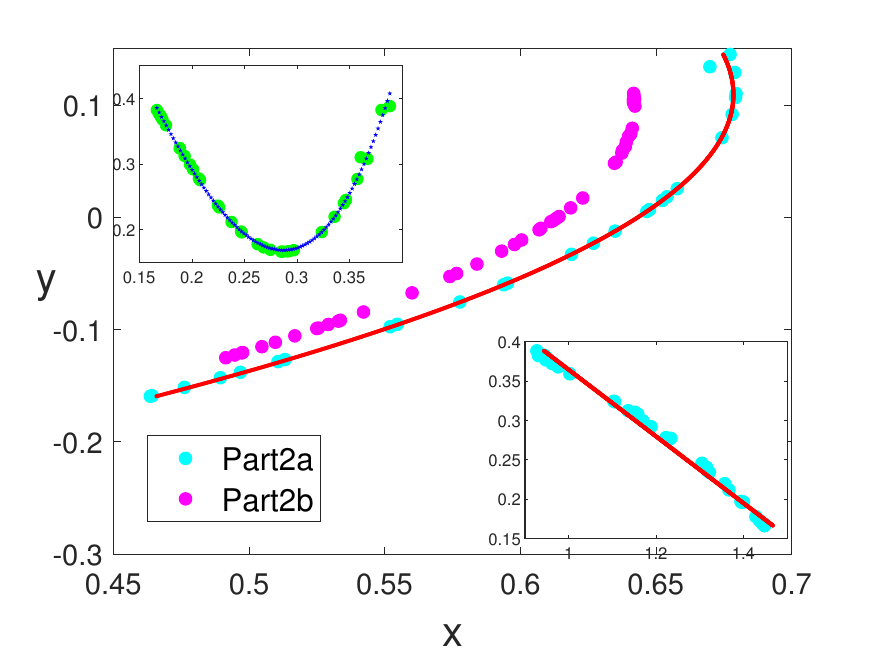}
				\caption{}
			\end{subfigure}
		\end{minipage}
		\hfill
		\begin{minipage}[t]{0.48\linewidth}
			\centering
			\begin{subfigure}[t]{\textwidth}
				\centering
				\includegraphics[width=8.5cm]{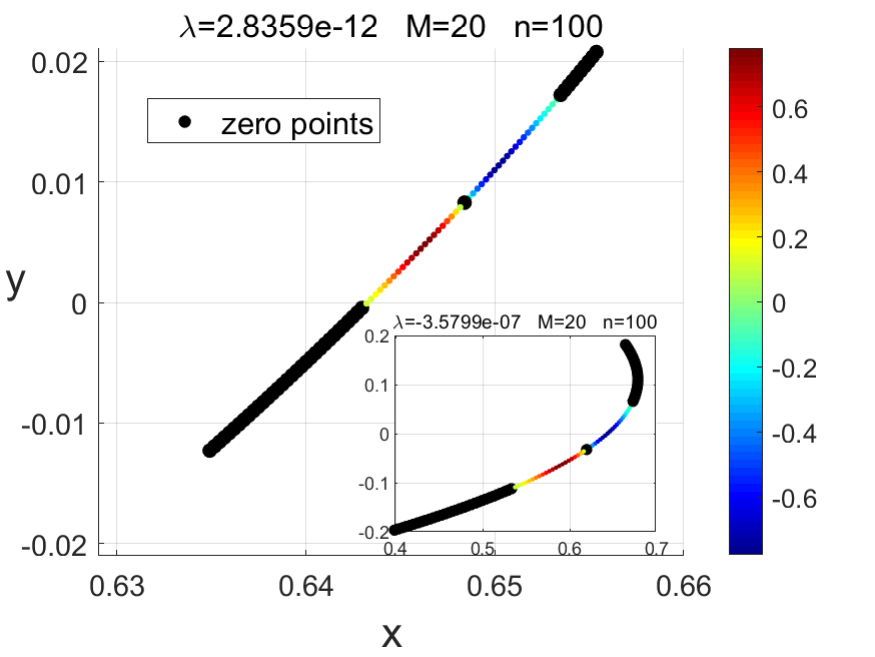}
				\caption{}
			\end{subfigure}
		\end{minipage}
		\caption{
			(a)  The interpolation for state points for Part2a via setting $n=100$
			(b) The VLEZ via KA method for Part2a when $M=20$. 
		}
		\label{fig:symbol henonB appd}
	\end{figure*}
	
	\begin{figure*}[!htbp] 
		
		\begin{minipage}[t]{0.48\linewidth}
			\centering
			\begin{subfigure}[t]{\textwidth}
				\centering
				\includegraphics[width=8.5cm]{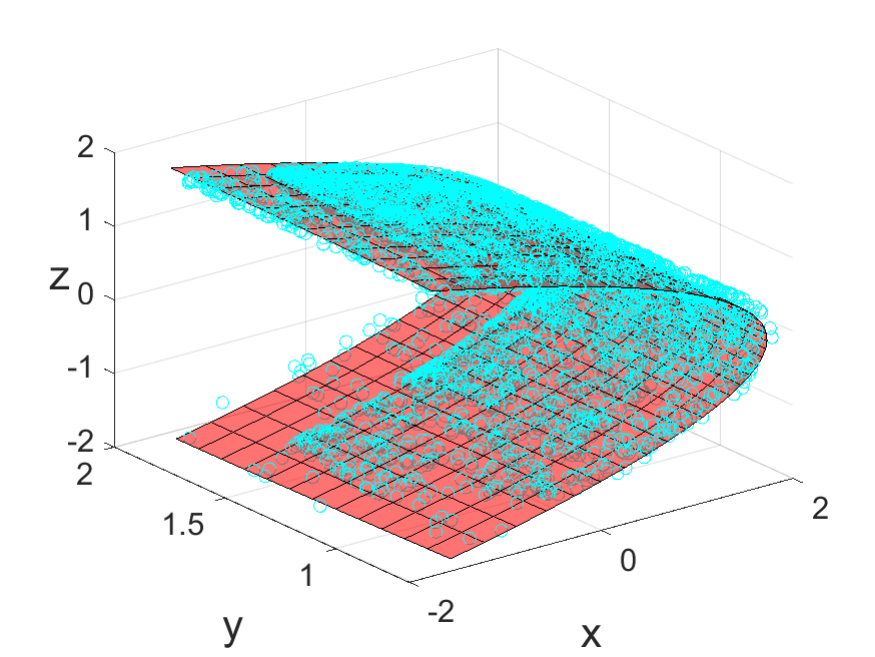}
				\caption{}
			\end{subfigure}
		\end{minipage}
		\hfill
		\begin{minipage}[t]{0.48\linewidth}
			\centering
			\begin{subfigure}[t]{\textwidth}
				\centering
				\includegraphics[width=8.5cm]{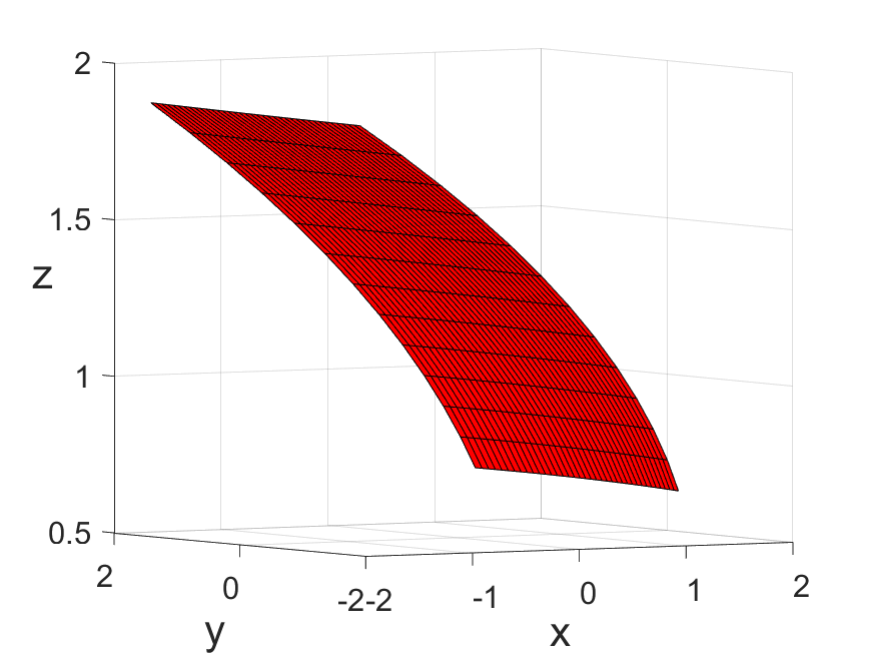}
				\caption{}
			\end{subfigure}
		\end{minipage}
		
		\caption{The generalized affine transformation for compact subregion
			(a)The fitting surface of second symbolic region
			(b)The fitting surface of one-step evolved second symbolic region 
		}
		\label{fig:three_local}
	\end{figure*}

	We determine whether the original region can be a curved region, with the constraint that the points of the original region and the fitted curve points have a coefficient of determination $R2\geq 0.9$. The formula for calculating the coefficient of determination is 
	\begin{equation}
		R^{2} = 1 -\frac{SSR}{SST} 
		\label{eq_R2}
	\end{equation}
	where SSR (Sum of Squares Residual) is the residual sum of squares, representing the difference between the values of the fitted curve and the actual state points. We refer to the fitting curve from original region as Curve O while refer that from evolved region as Curve P.
	
	First, based on the curve equation, we uniformly insert a sufficient number of equally spaced independent variables $\mathbf{x_\text{f}}$. By calculating the total arc length $l$ of the curve in the given direction, we can subsequently divide this arc length into equal segments $\Delta l$. This gives us infinitesimal curve elements of equal arc length. Using interpolation
	
	\begin{equation}
		\frac{\mathbf{x_\text{f\_i}}}{\mathbf{x_\text{f\_i}}}  = \frac{\Delta l}{l}, i=x,y,z,\ldots
		\label{eq_interp1}
	\end{equation}
	where $i$ means the direction of the independent variable $\mathbf{x_\text{f}}$.
	We can correspond the coordinates of $\mathbf{x_\text{f\_i}}$ on the curve with the arc length, and furthermore obtain the coordinates of $\mathbf{x_\text{f}}$ as the points with equal arc length as yellow points shown in the Fig.\ref{fig:henon1_whole}(c). When the numbers $\Delta l$ for the Curve O and Curve P are both $n-1$, we can use two sets of points $\mathbf{x_\text{f}}$ as the state points $\mathbf{x_\text{f\_o}}$ and evolved state points $\mathbf{x_\text{f\_aff}}$ respectively, and then correspond them according to the continuous direction of these two curves.
	Next, we find the proximate points $\mathbf{x_\text{f\_aff}}$ on original regions corresponding to the $\mathbf{x}$ on the Curve O.
	Then we find the corresponding evolved points of $\mathbf{x}$ on evolved regions,  $\mathbf{x_\text{p}}$. Then we map the $\mathbf{x_\text{p}}$ into the Curve P based on formula and  obtain $\mathbf{x_\text{f\_p}}$. 
	Similarly, when setting the number of $\Delta l$ for the Curve P is $M$, we can obtain the  coordinates of grid points $\mathbf{x^{*}}$. The  $\Delta l$ is consider as the radius width of the wave packet $d$ in that the arc length is approximately equal to the chord length, the distance between adjacent grid points, at a small scale. 
	We map $\mathbf{x_\text{f\_aff}}$, $\mathbf{x_\text{f\_p}}$, $\mathbf{x^{*}}$ and $d$ into the two-dimensional Gaussian basis function Eq.\eqref{gauss} to construct $K$ and $L$. 
	Then we can perform KA method to obtain the partition points on Curve P and obtain the VLEZ.
	We return the independent variable $\mathbf{x_\text{aff}}$ of VLEZ into  $\mathbf{x_\text{f\_o}}$ on the Curve O, and then further back into $\mathbf{x}$ on the the original region. Since the function value remains unchanged, we can obtain the VLEZ of the original region, thereby determining its symbolic partition boundary.
	We call the process of transfer $\mathbf{x}$ into $\mathbf{x_\text{aff}}$ as the Generalized Affine Transformation in that the curve is seen as straight line at a small scale. In other words, the transformation is extremely similar with linear transformation at a small scale. 
	The supplement for explanation for GKA method on two-dimensional chaotic map as presented in Fig.\ref{fig:symbol henonA appd} and Fig.\ref{fig:symbol henonB appd}.
	We can generalize this transformation to higher dimensional curving regions. However, we must to fitting curved surface if the localized region is too compact to separate into any curved-type regions just as case in section\ref{threedim}. Similarly, we take the similar steps except choosing the Gaussian functions in that we cannot ensure the grid points are uniform. Therefore, we continue to use the rectangle basis functions step by step and obtain refined symbolic boundary.
	The supplement for explanation for GKA method on three-dimensional chaotic map are presented in Fig.\ref{fig:three_local}.

	\bibliographystyle{unsrt}
	\bibliography{refsXXX}

\begin{thebibliography}{10}

\bibitem{1938symbol}
Morse M{,}~Hedlund GA.
\newblock Symbolic dynamics.
\newblock {\em American Journal of Mathematics}, 60(4):815--866, 1938.
\newblock doi: {\color{blue} \href{https://doi.org/10.2307/2371264}
  {https://doi.org/10.2307/2371264}}.

\bibitem{1995An}
Lind D{,}~Marcus M.
\newblock {\em An Introduction to Symbolic Dynamics and Coding}.
\newblock Cambridge University Press, 1995.

\bibitem{1988chaos}
Hao BL{,}~Zheng WM.
\newblock {\em Applied Symbolic Dynamics and Chaos}.
\newblock World Scientific, 2nd edition, 2018.

\bibitem{1987Recurrence}
Eckmann JP{,} Kamphorst SO{,}~Ruelle D.
\newblock Recurrence plots of dynamical systems.
\newblock {\em Europhys Lett}, 4, 1987.
\newblock doi: {\color{blue}\href{https://doi.org/10.1209/0295-5075/4/9/004}
  {https://doi.org/10.1209/0295-5075/4/9/004}}.

\bibitem{MARWAN2007237}
Norbert M{,} Carmen MR{,} Marco~T{,} et~al.
\newblock Recurrence plots for the analysis of complex systems.
\newblock {\em Physics reports}, 438(5-6):237--329, 2007.
\newblock doi:
  {\color{blue}\href{https://doi.org/10.1016/j.physrep.2006.11.001}
  {https://doi.org/10.1016/j.physrep.2006.11.001}}.

\bibitem{2017Elements}
Cover TM.
\newblock {\em Elements of Information Theory}.
\newblock Wiley Series in Telecommunications and Signal Processing, 2017.

\bibitem{1985Ergodic}
Eckmann JP{,}~Ruelle D.
\newblock Ergodic theory of chaos and strange attractors.
\newblock {\em Rev Mod Phys}, 57(3):617, 1985.

\bibitem{2009Complex}
Yuan LG{,} Nie DX{,}~Fu XC.
\newblock Complex orbits in a second-order digital filter with sinusoidal
  response.
\newblock {\em Chaos Soliton Fract}, 40(4):1660--1667, 2009.
\newblock doi: {\color{blue} \href{https://doi.org/10.1016/j.chaos.2007.09.048}
  {https://doi.org/10.1016/j.chaos.2007.09.048}}.

\bibitem{DONG20221}
Q{,} et~al Dong CW{,} Liu HH{,}~Jie.
\newblock Topological classification of periodic orbits in the generalized
  lorenz-type system with diverse symbolic dynamics.
\newblock {\em Chaos Soliton Fract}, 154:111686, 2022.
\newblock doi: {\color{blue}\href{https://doi.org/10.1016/j.chaos.2021.111686}
  {https://doi.org/10.1016/j.chaos.2021.111686}}.

\bibitem{KARAMANOS1999}
Karamanos K{,}~Nicolis G.
\newblock Symbolic dynamics and entropy analysis of feigenbaum limit sets.
\newblock {\em Chaos Soliton Fract}, 10(7):1135--1150, 1999.
\newblock doi:
  {\color{blue}\href{https://doi.org/10.1016/S0960-0779(98)00095-2}
  {https://doi.org/10.1016/S0960-0779(98)00095-2}}.

\bibitem{2008Symbolic}
Caneco A{,} Grácio C{,}~Rocha JL.
\newblock Symbolic dynamics and chaotic synchronization in coupled duffing
  oscillators.
\newblock {\em J Nonlinear Math Phys}, 15:102--110, 2008.
\newblock doi: {\color{blue}\href{https://doi.org/10.2991/jnmp.2008.15.s3.11}
  {https://doi.org/10.2991/jnmp.2008.15.s3.11}}.

\bibitem{X1995Symbol}
Tang XZ{,} Trac ER{,} Boozer~AD{,} et~al.
\newblock Symbol sequence statistics in noisy chaotic signal reconstruction.
\newblock {\em Phys Rev E}, 51(5):3871–3889, 1995.
\newblock doi: {\color{blue} \href{https://doi.org/10.1103/PhysRevE.51.3871}
  {https://doi.org/10.1103/PhysRevE.51.3871}}.

\bibitem{Hayes1993CommunicatingWC}
Hayes S{,} Grebogi C{,}~Ott E.
\newblock Communicating with chaos.
\newblock {\em Phys Rev Lett}, 70 20:3031--3034, 1993.
\newblock doi: {\color{blue}\href{https://doi.org/10.1103/PhysRevLett.70.3031}
  {https://doi.org/10.1103/PhysRevLett.70.3031}}.

\bibitem{1994Experimental}
Hayes S{,} Grebogi C{,} Ott~E{,} et~al.
\newblock Experimental control of chaos for communication.
\newblock {\em Phys Rev Lett}, 73(13):1781, 1994.
\newblock doi: {\color{blue}\href{https://doi.org/10.1103/PhysRevLett.73.1781}
  {https://doi.org/10.1103/PhysRevLett.73.1781}}.

\bibitem{2001Symbolic}
Edwards R{,} Siegelmann HT{,} Aziza~K{,} et~al.
\newblock Symbolic dynamics and computation in model gene networks.
\newblock {\em Chaos}, 11(1):160--160, 2001.
\newblock doi: {\color{blue}\href{https://doi.org/10.1063/1.1336498}
  {https://doi.org/10.1063/1.1336498}}.

\bibitem{2010chaos}
Predrag C{,} Roberto A{,} Ronnie~M{,} et~al.
\newblock Chaos: classical and quantum.
\newblock {\em ChaosBook. org (Niels Bohr Institute, Copenhagen 2005)}, 69:25,
  2005.

\bibitem{1985Generating}
Grassberger P{,}~Kantz H.
\newblock Generating partitions for the dissipative hénon map.
\newblock {\em Physics Lett A}, 113(5):235--238, 1985.
\newblock doi: {\color{blue}\href{https://doi.org/10.1016/0375-9601(85)90016-7}
  {https://doi.org/10.1016/0375-9601(85)90016-7}}.

\bibitem{1989On}
Grassberger P{,} Kantz H{,}~Moenig U.
\newblock On the symbolic dynamics of the henon map.
\newblock {\em J Phys A}, 22(22):5217, 1989.
\newblock doi: {\color{blue}\href{https://doi.org/10.1088/0305-4470/22/24/011}
  {https://doi.org/10.1088/0305-4470/22/24/011}}.

\bibitem{1995A}
Christiansen F{,}~Politi A.
\newblock A generating partition for the standard map.
\newblock {\em Phys Rev E}, 51(5):R3811, 1995.
\newblock doi: {\color{blue} \href{https://doi.org/10.1103/PhysRevE.51.R3811}
  {https://doi.org/10.1103/PhysRevE.51.R3811}}.

\bibitem{2021Symbolic}
Chai MS{,}~Lan YH.
\newblock Symbolic partition in chaotic maps.
\newblock {\em Chaos}, 2021.
\newblock doi: {\color{blue} \href{https://doi.org/10.1063/5.0042705}
  {https://doi.org/10.1063/5.0042705}}.

\bibitem{2002chaos}
Ott E.
\newblock {\em Chaos in Dynamical Systems}.
\newblock Cambridge University Press, 2002.

\bibitem{1991Model}
Flepp L{,} Holzner R{,} Brun~E{,} et~al.
\newblock Model identification by periodic-orbit analysis for nmr-laser chaos.
\newblock {\em Phys Lett A}, 153(17):2244--2247, 1991.
\newblock doi: {\color{blue} \href{https://doi.org/10.1103/PhysRevLett.67.2244}
  {https://doi.org/10.1103/PhysRevLett.67.2244}}.

\bibitem{1994Progress}
Badii R{,} Brun E{,}~Finardi M.
\newblock Progress in the analysis of experimental chaos through periodic
  orbits.
\newblock {\em Rev Mod Phys}, 66(4):1389--1415, 1994.
\newblock doi: {\color{blue} \href{https://doi.org/10.1103/RevModPhys.66.1389}
  {https://doi.org/10.1103/RevModPhys.66.1389}}.

\bibitem{2000Estimating}
Davidchack R{,} Lai YC{,} Brun~E{,} et~al.
\newblock Estimating generating partitions of chaotic systems by unstable
  periodic orbits.
\newblock {\em Phys Rev E}, 61(2):1353, 2000.
\newblock doi: {\color{blue} \href{https://doi.org/10.1103/PhysRevE.61.1353}
  {https://doi.org/10.1103/PhysRevE.61.1353}}.

\bibitem{1994Combining}
Lefranc M{,} Glorieux P{,} Papoff~F{,} et~al.
\newblock Combining topological analysis and symbolic dynamics to describe a
  strange attractor and its crises.
\newblock {\em Phys Rev Lett}, 73(10):1364--1367, 1994.
\newblock doi: {\color{blue} \href{https://doi.org/10.1103/PhysRevLett.73.1364}
  {https://doi.org/10.1103/PhysRevLett.73.1364}}.

\bibitem{1999From}
Plumecoq J{,}~Lefranc M.
\newblock From template analysis to generating partitions i: Periodic orbits,
  knots and symbolic encodings.
\newblock {\em Physica D}, 144(3):231--258, 1999.
\newblock doi: {\color{blue}
  \href{https://doi.org/10.1016/S0167-2789(00)00082-8}
  {https://doi.org/10.1016/S0167-2789(00)00082-8}}.

\bibitem{2000From}
Plumecoq J{,}~Lefranc M.
\newblock From template analysis to generating partitions ii: Characterization
  of the symbolic encodings.
\newblock {\em Physica D}, 144(3):259--278, 2000.
\newblock doi: {\color{blue}
  \href{https://doi.org/10.1016/S0167-2789(00)00083-X}
  {https://doi.org/10.1016/S0167-2789(00)00083-X}}.

\bibitem{1982Symbolic}
Crutchfield JP{,}~Packard NH.
\newblock Symbolic dynamics of one-dimensional maps: Entropies, finite
  precision, and noise.
\newblock {\em Int J Theor Phys}, 21(6):433--466, 1982.
\newblock doi: {\color{blue} \href{https://doi.org/10.1007/BF02650178}
  {https://doi.org/10.1007/BF02650178}}.

\bibitem{1983Symbolic}
Crutchfield JP{,}~Packard NH.
\newblock Symbolic dynamics of noisy chaos.
\newblock {\em Physica D}, 7(1-3):201--223, 1983.
\newblock doi: {\color{blue}
  \href{https://doi.org/10.1016/0167-2789(83)90127-6}
  {https://doi.org/10.1016/0167-2789(83)90127-6}}.

\bibitem{2003Estimating}
Kennel MB{,}~Buhl M.
\newblock Estimating good discrete partitions from observed data: symbolic
  false nearest neighbors.
\newblock {\em AIP Conf Proc}, 91:084102, 2003.
\newblock doi: {\color{blue}
  \href{https://doi.org/10.1103/PhysRevLett.91.084102}
  {https://doi.org/10.1103/PhysRevLett.91.084102}}.

\bibitem{2005Statistically}
Buhl M{,}~Kennel MB.
\newblock Statistically relaxing to generating partitions for observed
  time-series data.
\newblock {\em Phys Rev E}, 71(4-2):046213, 2005.
\newblock doi: {\color{blue} \href{https://doi.org/10.1103/PhysRevE.71.046213}
  {https://doi.org/10.1103/PhysRevE.71.046213}}.

\bibitem{2004Estimating}
Hirata Y{,} Judd K{,}~Kilminster D.
\newblock Estimating a generating partition from observed time series: Symbolic
  shadowing.
\newblock {\em Phys Rev E}, 70(1-2):016215, 2004.
\newblock doi: {\color{blue} \href{https://doi.org/10.1103/PhysRevE.70.016215}
  {https://doi.org/10.1103/PhysRevE.70.016215}}.

\bibitem{2013Estimating}
Hirata Y{,}~Aihara K.
\newblock Estimating optimal partitions for stochastic complex systems.
\newblock {\em Eur Phys J Spec Topics}, 222(2):303--315, 2013.
\newblock doi: {\color{blue} \href{https://doi.org/10.1140/epjst/e2013-01843-x}
  {https://doi.org/10.1140/epjst/e2013-01843-x}}.

\bibitem{2018Empirical}
Patil NS{,}~Cusumano JP.
\newblock Empirical generating partitions of driven oscillators using optimized
  symbolic shadowing.
\newblock {\em Phys Rev E}, 98(3), 2018.
\newblock doi: {\color{blue} \href{https://doi.org/10.1103/PhysRevE.98.032211}
  {https://doi.org/10.1103/PhysRevE.98.032211}}.

\bibitem{2020The}
Patil NS{,}~Cusumano JP.
\newblock The high forecasting complexity of stochastically perturbed periodic
  orbits limits the ability to distinguish them from chaos.
\newblock {\em Nonlinear Dynam}, 102(1):1--16, 2020.
\newblock doi: {\color{blue} \href{https://doi.org/10.1007/s11071-020-05920-z}
  {https://doi.org/10.1007/s11071-020-05920-z}}.

\bibitem{1997Finding}
Allie S{,}~Mees A.
\newblock Finding periodic points from short time series.
\newblock {\em Phys Review E}, 56(1):346--350, 1997.
\newblock doi: {\color{blue} \href{https://doi.org/10.1103/PhysRevE.56.346}
  {https://doi.org/10.1103/PhysRevE.56.346}}.

\bibitem{1997Detecting}
Schmelcher P{,}~Diakonos FK.
\newblock Detecting unstable periodic orbits of chaotic dynamical systems.
\newblock {\em Phys Rev Lett}, 78(25):4733--4736, 1997.
\newblock doi: {\color{blue} \href{https://doi.org/10.1103/PhysRevLett.78.4733}
  {https://doi.org/10.1103/PhysRevLett.78.4733}}.

\bibitem{1931Hamiltonian}
Koopman BO.
\newblock Hamiltonian systems and transformation in hilbert space.
\newblock {\em Proc Natl Acad Sci}, 17(5):315--318, 1931.
\newblock doi: {\color{blue} \href{https://doi.org/10.1073/pnas.17.5.315}
  {https://doi.org/10.1073/pnas.17.5.315}}.

\bibitem{2010Dynamic}
Schmid PJ.
\newblock Dynamic mode decomposition of numerical and experimental data.
\newblock {\em J Fluid Mech}, 656(10):5--28, 2010.
\newblock doi: {\color{blue} \href{https://doi.org/10.1017/S0022112010001217}
  {https://doi.org/10.1017/S0022112010001217}}.

\bibitem{Williams2014ADA}
Williams MO{,} Kevrekidis IG{,}~Rowley CW.
\newblock A data–driven approximation of the koopman operator: Extending
  dynamic mode decomposition.
\newblock {\em J Nonlinear Sci}, 25:1307 -- 1346, 2014.
\newblock doi: {\color{blue} \href{https://doi.org/10.1007/s00332-015-9258-5}
  {https://doi.org/10.1007/s00332-015-9258-5}}.

\bibitem{2009Spectral}
Rowley CW{,} Mezi{\'c} I{,} Bagheri~S{,} et~al.
\newblock Spectral analysis of nonlinear flows.
\newblock {\em J Fluid Mech}, 2009.
\newblock doi: {\color{blue} \href{https://doi.org/10.1017/S0022112009992059}
  {https://doi.org/10.1017/S0022112009992059}}.

\bibitem{Bagheri2013}
Bagheri S.
\newblock Koopman-mode decomposition of the cylinder wake.
\newblock {\em J Fluid Mech}, 726:596–623, 2013.
\newblock doi: {\color{blue} \href{https://doi.org/10.1017/jfm.2013.249}
  {https://doi.org/10.1017/jfm.2013.249}}.

\bibitem{2011Applications}
Schmid PJ{,} Li L{,}~Juniper MP.
\newblock Applications of the dynamic mode decomposition.
\newblock {\em Theor Comput Fluid Dyn}, 25(1-4):249--259, 2011.
\newblock doi: {\color{blue} \href{https://doi.org/10.1007/s00162-010-0203-9}
  {https://doi.org/10.1007/s00162-010-0203-9}}.

\bibitem{Eisenhower2010DECOMPOSINGBS}
Eisenhower B{,} Maile T{,} Fischer~M{,} et~al.
\newblock Decomposing building system data for model validation and analysis
  using the koopman operator.
\newblock In {\em Proceedings of SimBuild Conference 2010}. IPBSA, 2010.

\bibitem{2012CREATING}
B{,} Mezi{\'c}~I Georgescu M{,}~Eisenhower.
\newblock Creating zoning approximations to building energy models using the
  koopman operator.
\newblock {\em Proceedings of SimBuild}, 5(1):40--47, 2012.

\bibitem{2011Nonlinear}
Susuki Y{,}~Mezi{\'c} I.
\newblock Nonlinear koopman modes and coherency identification of coupled swing
  dynamics.
\newblock {\em IEEE T Power Syst}, 26(4):1894--1904, 2011.
\newblock doi: {\color{blue} \href{https://doi.org/10.1109/TPWRS.2010.2103369}
  {https://doi.org/10.1109/TPWRS.2010.2103369}}.

\bibitem{2012Nonlinear}
Susuki Y{,}~Mezi{\'c} I.
\newblock Nonlinear koopman modes and a precursor to power system swing
  instabilities.
\newblock {\em IEEE T Power Syst}, 27(3):1182--1191, 2012.
\newblock doi: {\color{blue} \href{https://doi.org/10.1109/TPWRS.2012.2183625}
  {https://doi.org/10.1109/TPWRS.2012.2183625}}.

\bibitem{Mezi2020KoopmanOG}
Mezi{\'c} I.
\newblock Koopman operator, geometry, and learning.
\newblock {\em Dyn Syst}, 2020.

\bibitem{2022Phase}
Zhang C{,} Li H{,}~Lan Y.
\newblock Phase space partition with koopman analysis.
\newblock {\em Chaos}, 32(6), 2022.
\newblock doi: {\color{blue} \href{https://doi.org/10.1063/5.0079812}
  {https://doi.org/10.1063/5.0079812}}.

\bibitem{generalized}
Baier G{,}~Klein M.
\newblock Maximum hyperchaos in generalized hénon maps.
\newblock {\em Phys Lett A}, 151(6):281--284, 1990.
\newblock doi: {\color{blue}
  \href{https://doi.org/10.1016/0375-9601(90)90283-T}
  {https://doi.org/10.1016/0375-9601(90)90283-T}}.

\bibitem{M1976A}
H{\'e}non M.
\newblock A two-dimensional mapping with a strange attractor.
\newblock {\em Commun math Phys}, 50(1), 1976.
\newblock doi: {\color{blue} \href{https://doi.org/10.1007/BF01608556}
  {https://doi.org/10.1007/BF01608556}}.

\bibitem{1997Structure}
Jaeger L{,}~Kantz H.
\newblock Structure of generating partitions for two-dimensional maps.
\newblock {\em J Phys A}, 3097(16):567--576, 1997.

\bibitem{hensethesis}
Hansen KT.
\newblock {\em Symbolic dynamics in chaotic systems}.
\newblock PhD thesis, University of Oslo, 1993.

\bibitem{1991Homoclinic}
Giovannini F{,}~Politi A.
\newblock Homoclinic tangencies, generating partitions and curvature of
  invariant manifolds.
\newblock {\em J Phys A}, 24(8):1837, 1991.
\newblock doi: {\color{blue} \href{https://doi.org/10.1088/0305-4470/24/8/024}
  {https://doi.org/10.1088/0305-4470/24/8/024}}.

\end{thebibliography}

\end{document}